\documentclass[english, 12pt, renew]{article}
\usepackage{tikz}
\usepackage{verbatim}
\usetikzlibrary{decorations.pathreplacing,arrows.meta,patterns,shapes, calligraphy}
\usepackage{mathtools}
\usepackage{ae,aecompl}
\usepackage[T1]{fontenc}
\usepackage[utf8]{inputenc}
\usepackage[margin = 1in]{geometry}
\usepackage{geometry}
\usepackage[nobiblatex]{xurl}

\makeatletter
\newcommand*{\addFileDependency}[1]{
  \typeout{(#1)}
  \@addtofilelist{#1}
  \IfFileExists{#1}{}{\typeout{No file #1.}}
}
\makeatother

\usepackage{lscape} 
\usepackage{hhline}
\usepackage{makecell, caption, booktabs}
\usepackage{siunitx}
\usepackage{color}
\usepackage{amsmath}
\usepackage{amsthm}
\usepackage{amssymb}
\usepackage{wasysym}
\usepackage{amsfonts}
\usepackage{latexsym}
\usepackage{tabularx}
\usepackage[authoryear]{natbib}
\usepackage{blindtext}
\usepackage{graphicx}
\usepackage{subcaption}
\usepackage[utf8]{inputenc}
\usepackage[export]{adjustbox}
\usepackage{wrapfig}
\usepackage{hyperref}
\usepackage{breakurl}
\usepackage{amssymb,amsmath}
\usepackage{graphicx}
\usepackage{fancyhdr}
\usepackage{fancybox}
\usepackage{shadow}
\usepackage{empheq}
\usepackage{titlesec}
\usepackage{titletoc}
\usepackage{soul, color}
\usepackage{float}
\usepackage{shorttoc}
\usepackage{xcolor}
\usepackage{fancybox}
\usepackage{array}
\usepackage{tabularx}
\usepackage{booktabs}
\usepackage{rotating}
\usepackage{bigstrut}
\usepackage{aeguill}
\usepackage{subcaption}
\usepackage{authblk}

\usepackage{breakurl}
  \theoremstyle{definition}
  
  \theoremstyle{plain}
  
\theoremstyle{plain}

  \theoremstyle{plain}
  
  \theoremstyle{remark}
  
  \theoremstyle{plain}
  
 \theoremstyle{definition}
  
   \theoremstyle{definition}
  
  \theoremstyle{plain}
  
  \theoremstyle{plain}
  \newtheorem{proposition}{\protect\propositionname}
  \theoremstyle{definition}

  \theoremstyle{remark}

\newcommand*\ab{.4}
  \tikzset{
    net node/.style = {circle, minimum width=2*\ab cm, inner sep=0pt, outer sep=0pt, ball color=blue!50!cyan},
    net root node/.style = {net node, minimum width=3*\ab cm},
    net connect/.style = {line width=1pt, draw=blue!50!cyan!25!black},
  }

\makeatother

  \providecommand{\axiomname}{Axiom}
  \providecommand{\definitionname}{Definition}
  \providecommand{\examplename}{Example}
  \providecommand{\lemmaname}{Lemma}
  \providecommand{\remarkname}{Remark}
\providecommand{\corollaryname}{Corollary}
\providecommand{\theoremname}{Theorem}
\providecommand{\notationname}{Notation}
\providecommand{\resultname}{Result}
\providecommand{\propositionname}{Proposition}
\providecommand{\assumptionname}{Assumption}
 \providecommand{\claimname}{Claim}
\title{\large{\textbf{
Optimally Targeting Interventions in Networks during a Pandemic: Theory and Evidence from the Networks of Nursing Homes in the United States}\thanks{We are grateful to Editor Gregory Ponthiere and three anonymous referees for insightful and constructive comments and suggestions that have helped improve our study. We are also grateful for the remarks by Randall Monty on the early version of the paper, and we acknowledge valuable discussions and suggestions from several seminars and conference participants. Special thanks are due to Lidia Liu for excellent research assistance.  Pongou acknowledges financial support from the SSHRC's Partnership Engage Grants COVID-19 Special Initiative. Pongou: University of Ottawa and Harvard T.H. Chan School of Public Health, rpongou@uottawa.ca or rpongou@hsph.harvard.edu; Tchuente: University of Kent, g.tchuente@kent.ac.uk; Tondji:  The University of Texas Rio Grande Valley, jeanbaptiste.tondji@utrgv.edu.}} \vspace{-.5em}}
\author{Roland Pongou  \  \ Guy Tchuente \ \ Jean-Baptiste Tondji \vspace{-1em}}

\begin{document}
\maketitle
\thispagestyle{empty}
\vspace{-3em}

\begin{abstract}
\footnotesize{This study develops an economic model for a social planner who \textit{prioritizes} health over short-term wealth accumulation during a pandemic. Agents are connected through a weighted undirected network of contacts, and the planner's objective is to determine the policy that contains the spread of infection below a \textit{tolerable} incidence level, and that maximizes the present discounted value of real income, in that order of priority.  The optimal unique  policy depends both on the configuration of the contact network and the tolerable infection incidence. Comparative statics analyses are conducted: (i) they reveal the tradeoff between the economic cost of the pandemic and the infection incidence allowed; and (ii) they suggest a correlation between different measures of network centrality and individual lockdown probability with the correlation increasing with the \textit{tolerable} infection incidence level. Using unique data on the networks of nursing and long-term homes in the U.S., we calibrate our model at the state level and estimate the tolerable COVID-19 infection incidence level. We find that \textit{laissez-faire} (more tolerance to the virus spread) pandemic policy is associated with an increased number of deaths in nursing homes and higher state GDP growth. In terms of the death count, \textit{laissez-faire} is more harmful to nursing homes than more peripheral in the networks, those located in deprived counties, and those who work for a profit. We also find that U.S. states with a Republican governor have a higher level of \textit{tolerable} incidence, but policies tend to converge with high death count.
}
\end{abstract}

\clearpage
\pagenumbering{arabic} 

\section{Introduction}

\ \ \  In this study, we develop an economic theory that addresses the problem of finding the optimal lockdown policy during a pandemic that spreads through networks of physical contacts, for a social planner who \textit{prioritizes} health over short-term wealth accumulation. We apply this theory to the current coronavirus pandemic, and uncover new results on the effects of network configuration, network centrality, and health policies. Using unique data on the networks of nursing and long-term care homes in the United States, we calibrate our model and empirically validate its key predictions. 

The application of our model to the spread of the novel coronavirus disease (COVID-19) is timely and fitting. COVID-19 has affected millions of individuals and claimed many lives globally. Two common strategies to counteracting the epidemic spread are eradication and suppression. Governments around the world are implementing a variety of strategies to contain this pandemic: prescriptions of social distancing and hygiene measures, extensive production of personal protection equipment (PPE), expansion of testing and hospital capacities, use of face masks and contact tracing, and vaccines.  These objectives resulted in the enforcement of social distancing policies that led to the cumulative lockdown of over half of the world population \citep{statista1}. While this approach for mitigating the contagion has shown some positive results, the associated economic costs are considerable \citep{kochanczyk2021pareto}. The Gross Domestic Product in both advanced and developing countries has decreased significantly as a result of the COVID-19 pandemic \citep{IMF2020}. In a situation where only a small fraction of jobs can be done remotely \citep{Dingel2020}, such mitigation measures may not be economically viable in the long run. The significant economic and social costs implied by quasi-complete lockdown have forced governments and policymakers to think about less costly alternatives that might consist of imposing quarantine measures only on certain individuals while letting others go back to work. The natural question that arises is: how do we design optimally targeted lockdown policies that account for social network structure, and how do these policies affect health and economic dynamics?

We address this question in a society that \textit{prioritizes} health over short-term wealth accumulation.\footnote{We view our main question as a planning problem, and our assumption that health is prioritized over short-term economic gains is consistent with several recent observations \citep{John2020, Heap2020, Aiello2020, baccini2021explaining}. For instance, \citeauthor{Stiglitz2020} wrote that: ``There can be no economic recovery until the virus is contained, so addressing the health emergency is the top priority for policymakers." \citep[n.d]{Stiglitz2020}.} We describe our model and state the social planner's objective problem: 

\begin{enumerate}
\item[A:] Agents (including individuals and social infrastructures) are connected through a weighted undirected network of physical contacts through which the virus is likely to spread. The weighted nature of the contact network implies that the intensity of interactions between agents is not necessarily binary, and it is likely to vary across relationships.\footnote{Our model allows the interpretation of a contact network to be broad. At the micro-level, agents are individuals and the infrastructures (for example,  nursing homes) they interact with daily. At the macro-level, agents can be viewed as spatial entities such as countries or cities. Moreover, network configuration is arbitrary, making sense given that social structure varies  across societies. For instance, some societies are individualistic, while others are organized around extended family and ethnic networks \citep{deji2011gender, pongou2010economics}.}

\item[B:] At any point in time, an agent is in one of the following four compartments: susceptible, infected, recovered, or dead. At time zero, all agents are susceptible. Susceptible agents can become infected, while infected agents can recover or die. Susceptible, infected and recovered agents can all be sent into lockdown. The dynamics of infection, recovery, and death follow a model that generalizes the classical SIR model \citep{kermack1927contribution, kermack1932contributions} in two ways. First, whereas the classical model assumes a random \textit{matching technology}, our model assumes any arbitrary network structure, and agents can only be infected through connections in the prevailing contact network if they have not been sent to lockdown. Second, whereas the classical model assumes three compartments (susceptible, infected, recovered), our model incorporates an additional compartment (death), and the lockdown variable. The lockdown variable is a key choice variable for the social planner, who uses it to modify the structure of the prevailing social network in order to slow the spread of contagion. 

\item[C:] The social planner's objective is to determine the lockdown policy that contains the spread of the infection below a tolerable infection \emph{incidence} level, and that maximizes the present discounted value of real income (or alternatively, that minimizes the economic cost of the pandemic), in that order of priority. In other words, the social planner can allocate the ``work-from-home'' rights to achieve these goals. An appeal of this approach to the social planner's problem is that it does not force us to assign a precise monetary value to health or to life.\footnote{See \citet{pindyck2020covid} and \citet{bosi2021optimal} for the expression of a similar concern.} Rather, it allows some flexibility in how to design policies, with clear health and economic goals in mind. For instance, the social planner could set an infection incidence level that allows to keep the number of infected individuals below hospitals' capacities, or they could set an incidence level that is high enough if they prioritize short-term economic gains over health. 
\end{enumerate}

We apply our theoretical model to analyze: (1) the effect of network structure on the dynamics of optimal lockdown, infection, recovery, death, and economic costs; (2) the tradeoff between public health and the economic cost of the pandemic; and (3) how  different measures of network centrality affect the probability of being sent to lockdown. In order to solve the planner's problem, we first characterize the dynamics of infection, recovery, and death rates in our N-SIRD epidemiological model with lockdown, and then we obtain a unique solution under classical conditions. Proposition \ref{propdynamics} shows that the rates of infection, recovery and death at any given time is a function of the lockdown variable and the initial network of contacts that captures social structure. Furthermore, we analyze the N-SIRD model using stability theory of differential equation, and we derive the basic reproduction number, $R_0$, from the largest eigenvalue of the next-generation matrix \citep{diekmann1990definition, van2002reproduction}. Then, Proposition \ref{stability} provides conditions for the local stability of disease-free equilibria, and Proposition \ref{finalsizesolution} generates the final size of the pandemic, both in terms of the basic reproduction number \citep{diekmann2000mathematical, wallinga2006using, andreasen2011final}.

The planner's objective is achieved in two ordered steps. The first step consists of identifying the set of lockdown policies that contain the epidemic spread (as dictated by the solution of the N-SIRD above) below a tolerable infection incidence threshold, and the second step consists of choosing among those contagion-minimizing policies the one which maximizes the discounted stream of economic surpluses. Proposition \ref{prop-sp} shows that the planner's problem admits a unique solution, and this solution depends on both the infection incidence level tolerated by the planner and the prevailing network of physical contacts that characterizes the society.  Proposition \ref{prop-sp} also shows that the tolerable infection incidence level and the prevailing network of physical contacts determine the dynamics of infection, recovery, death, and economic costs.  

We develop several comparative statics analyses of our theoretical findings using simulations that rely on realistic data on COVID-19 transmission rates.\footnote{In our simulation analyses, we assume that networks are represented by binary adjacent matrices $A=(A_{ij})$, where $A_{ij}=1$ if agents $i$ and $j$ are connected, and $A_{ij}=0$, otherwise. In our empirical analysis in Section \ref{sec:empiricalApp}, the network of nursing homes in each United States (U.S.) state is not necessarily binary since $A_{ij}$ ranges from 1 to 832 contacts, where $i$ and $j$ represent two distinct nursing homes. Since our empirical findings appear to be consistent with the simulation results, we believe that the binary nature of the network structure does not affect the quality of our findings. In our study, although we choose the values of the parameters of the production function to match as possible the data from the U.S. nursing and long-term care homes \citep{chen2021nursing}, we select other parameters in the planning problem for illustration purposes. Thus, one should interpret the quantitative outcomes of our model with caution. Nevertheless, in the empirical application in Section \ref{sec:empiricalApp}, we choose the economic parameters to match each U.S. state's reality as possible.} Figure \ref{lock_dyn_sw} illustrates how the tolerable infection incidence level and the prevailing network of physical contacts determine the dynamics of lockdown (Figure \ref{incidenceandlockdown}), infection (Figure \ref{incidenceandinfection}), death (Figure \ref{incidenceanddeaths}), and economic losses (Figure  \ref{incidenceandcosts}). Our result shows how the tolerable infection incidence affects the lockdown dynamics as well as the economic cost of the pandemic.  We find that a higher tolerable incidence level results in lower lockdown rates and economic surplus loss. While this result illustrates the health-vs-wealth tradeoff the social planner faces, it does not prescribe any resolution because the planning decision depends on how society values population health over short-term economic gains. Second, we illustrate how lattice, small-world, random, and scale-free network structures affect optimal lockdown probabilities and the disease dynamics, respectively. Our simulation results in Figure \ref{lock_dyn_net} show that the cumulative proportion of the population sent into lockdown is always higher in random and small-world network configurations compared to lattice and scale-free structures. These lockdown policies translate to different epidemic and economic costs dynamics for each network. We extend our analysis to examine the potential impact of \textit{network density} (or the interconnections between agents in a network) in our N-SIRD model for a small-world network. Our simulations (Figure \ref{lock_dyn_den}) show that optimal lockdown probabilities increase with network density (Figure \ref{lockdownden}).\footnote{Our series of robustness checks conjecture that the simulation results with lattice, random, and scale-free networks are qualitatively consistent with those obtained with the small-world network.} Third, we illustrate how measures of network centrality affect optimal lockdown probabilities and the disease dynamics. In Table \ref{tab_central}, we provide the correlations between four network metrics---degree, eigenvector, betweenness, and closeness---and average lockdown probabilities in a small-world network. Our simulation results suggest that individuals that are more central in such a network are more likely to be sent into lockdown. We provide a discussion of the robustness of these findings in Section \ref{centralitylockdown}. Overall, our simulation results confirm the intuition that not \textit{all} agents should be into complete lockdown under the optimal policy \citep{Acemoglu2020, gollier2020cost, gollier2020pandemic,ipwsj, bosi2021optimal, chang2021mobility, farsalinos2021improved}. This planning decision is justified since the goal in the N-SIRD model is to disconnect the prevailing contact network while maintaining economic activities, contrary to a pure epidemiological model.

We calibrate our model and test some of its key predictions using unique data from the networks of nursing and long-term care homes in the United States (U.S.). The senior population in the U.S. accounts for a significant share of COVID-19 deaths \citep{CDCseniorcovid19, kff2020, conlen2021more}. The surge of COVID-19 cases and deaths in U.S. nursing and long-term care homes led the American federal government to instate a ban to nursing home visits on March 13, 2020. This restriction enabled researchers from the ``Protect Nursing Homes" project to construct a U.S. nursing homes network, using smartphone data \citep{chen2021nursing}.\footnote{We are grateful to \citet{chen2021nursing} and all team members of the Protect Nursing Homes project (\url{https://www.protectnursinghomes.org/}) to generously share the nursing network data.} We use this unique network data in conjunction with nursing homes and U.S. state level data to calibrate the N-SIRD model. In the calibration exercise, we consider each nursing home as a node in the transmission network. Two nursing homes are connected if the same smartphone signal is recorded in both homes' locations. The number of distinct signals recorded gives a weight to the link between two nursing homes. The data shows that the weight ranges from 1 to 832.  Data on nursing homes contain the number of COVID-19 cases and deaths. We also collect information from the Senior Living project to complement the COVID-19 fatalities in the network data. We retrieve U.S. state level data  from the Bureau of Labor Statistics.\footnote{We gathered additional data from the website: \url{https://www.seniorliving.org/nursing-homes/costs/} consulted on September 9, 2021. We obtained the calibration of the epidemiological parameters from Statista: \url{https://www.statista.com/}.}

The calibration allows us to estimate the value of the tolerable infection incidence level, for 26 U.S. states. Following the calibration, the tolerable infection incidence level ($\lambda$) is estimated using a simulated minimum-distance estimator \citep{gertler1992quality, smith1993estimating, forneron2018abc}. The tolerable infection incidence level represents the tradeoff between population health and short-term economic gains. So, a higher value of tolerable infection incidence level describes a policy that tends more towards a ``laissez-faire" regime \citep{gollier2020cost}, indicating a planner's inclination to maximize short-term economic gains even if this results in more infections and deaths. We find that the tolerable infection incidence level varies significantly across U.S. states, making it possible to test some predictions of our N-SIRD epidemiological model.

Using regression-based analyses,  we find that a laissez-faire policy is associated with more deaths, consistent with our results from the simulations. We also find that a nursing home that is more central in the network experiences more COVID-19 deaths. However, laissez-faire reduces the death difference between \textit{central} nodes (or nursing homes with more connections) and those that are more peripheral. It follows that the vulnerability to the virus among peripheral nursing homes increases more under a laissez-faire regime relative to the expected increase under a more restrictive pandemic regime. We also find that laissez-faire increases COVID-19 fatalities of nursing homes located in more deprived counties. A laisser-faire policy also increases the vulnerability of for-profit nursing homes. Our regressions are robust, controlling for an array of variables at the nursing home and state levels, such as overall quality and county fixed-effects. In another empirical test of the N-SIRD model with lockdown, we investigate the relationship between the tolerable infection incidence level of COVID-19 and U.S. state GDP growth for the year 2020. We find that laissez-faire is associated with higher GDP growth, consistent with the model's prediction. The empirical estimation equation controls several factors, including the party affiliation and the gender of the U.S. state's governor. Interestingly, we find that the positive economic effect of laissez-faire is reduced for U.S. states whose governor is a Democrat. 

Finally, in an attempt to validate our calibration results using external information, we investigate the political origins of laissez-faire policies during the early period of the COVID-19 pandemic in the U.S.. We find that Republican governors are more likely to tolerate the pandemic. These results mirror the pro-market tendency of the republican party. However, these policy differences are minimized for U.S. states with a high death count in nursing homes. We also find that U.S. southern states are more prone to laissez-faire than states in other regions. These findings complement other studies showing an association between the political affiliation of a U.S. state governor and COVID-19 cases and deaths \citep{neelon2021associations, frankel2021virus, chen2021relationship}. These findings validate our calibration exercise, especially since the empirical framework doesn't use political variables.

Our work is related to several recent studies on virus spread in an undirected network. There is a substantial wealth of studies on the topic as surveyed by \citet{pastor2015epidemic}. This literature includes a class of mean-field models \citep{kephart1992directed, barabasi1999mean, green2006parameterization} and $N$-intertwined models via Markov theory in discrete time \citep{ganesh2005effect, wang2003epidemic} and continuous time \citep{van2008virus}. \citet{asavathiratham2001influence} and \citet{garetto2003modeling} review other general models for virus spread in networks based on Markov theory.\footnote{The $N$-intertwined models investigate the influence of the network topology on the spread of viruses, whose dynamics are described by different epidemiological models, including SIS and SIR. In these models, each node in the network is a Markov chain with the number of compartments corresponding to the selected epidemiological model. Independent Poisson processes describe the transition of the state of each node in the network. Then, scholars use mean-field approximations to capture the effect of neighbors on the total expected infection. Contrary to this growing literature, we examine the impact of network structure and reversible lockdown state in a SIRD model in continuous time with heterogeneous agents. One can use a Markov chain approach to extend our study.}

In the economics literature on COVID-19, the canonical SIR model has been generalized in several directions to address a variety of problems. Recent generalizations include among others,  \citet{Acemoglu2020} who propose a multi-risk SIR model, \citet{bethune2020covid} who study externalities of health interventions for infectious diseases in SIS and SIR models, \citet{karaivanov2020social} who examines the transmission of COVID-19 through a dynamic social-network model embedding the SIR model onto a graph of network contacts, \citet{alvarez2020simple}, \citet{bandyopadhyay2020learning}, \citet{federico2020taming}, \citet{eichenbaum2020macroeconomics}, \citet{gollier2020pandemic}, \cite{berger2020testing},  \citet{prem2020effect}, \citet{bisin2021spatial}, and \citet{ma2021intergenerational} who analyze optimal non-pharmaceutical controls in SIR models, \citet{chang2021mobility} who use Google mobility network and metapopulation susceptible–exposed–infectious–removed (SEIR) models to explain differences in COVID-19 fatalities and inform reopening in ten of the largest U.S. metropolitan areas, and \cite{Kuchler_covidfacebook2020} and \cite{Harris2020} who document the importance of social media networks (for example, Facebook) in the selection of targeted lockdown policies. While our model contains some of the ingredients of these other approaches, it differs in incorporating into the classical model two key elements, namely a lockdown variable and a weighted network of contacts that is not necessarily \textit{random}, and where agents are heterogeneous with respect to their degree of connections and individual characteristics. Most importantly, we introduce a \textit{lexicographic approach} to the planning problem, whereby the social planner's goal is to determine the lockdown policy that contains the spread of the infection below an acceptable incidence level, and that minimizes the economic cost of the pandemic, in that order of priority. An appeal of this approach is to help avoid the difficult problem of assigning a precise monetary value to health and life, which allows a great deal of flexibility and clarity in how to design optimal lockdown policies, with precise health and economic goals in mind. It also enables a transparent analysis of the tradeoff between public health and  short-term economic prosperity. 

In this study, our goal is to provide a dynamic lockdown rate that allows the planner to save lives at the minimum economic costs. To contain the contagion, \citet{bosi2021optimal} assume that the planner imposes a lockdown, which will stay constant over time. Contrary to \citet{bosi2021optimal}, our lockdown is reversible, and more in line with \citet{alvarez2020simple}, \citet{Acemoglu2020}, and \citet{gollier2020cost}.\footnote{We thank an anonymous referee for bringing this issue to our attention.} Assuming irreversible lockdown under a tractable epidemiological model enables the researcher to derive a closed-form solution while establishing the convexity of the problem with second-order conditions \citep{seierstad1986optimal, bosi2021optimal}. As in \citet{alvarez2020simple}, the interactions between our N-SIRD epidemiological model with a dynamic lockdown may make the problem non-convex. Therefore, we cannot verify the second-order condition given a lockdown profile candidate as an optimal solution. In other words, it would not be possible to prove that our optimal lockdown policy is indeed minimizing the economic costs of lockdown. Though we don't address the convexity issue, we follow \citet{alvarez2020simple} and \citet{Acemoglu2020} and use simulations to illustrate comparative analyses of our framework. In addition, we  provide an empirical application of the simulation results.

Our study is also connected to the economic literature on the design of interventions on networks and of the diffusion of knowledge or contagion via a network. \cite{ballester2006s}, and \cite{banerjee2013diffusion} examine the optimal targeting of the optimal player in a network, while \cite{galeotti2020targeting} analyze an optimal intervention of a social planner acting on individual incentives. The choice of the optimal lockdown in the social planner problem differentiates our model from these studies by being an intervention on the network structure. The lockdown operates to control the diffusion of the infection. Thus, it relates our work to those of \cite{young2009innovation}, and \cite{young2011dynamics} who investigate the diffusion of innovations through a network. Our work is also connected to the social learning dynamics as in \cite{buechel2015opinion}, and \cite{battiston2015boundedly} with the difference that the infection diffusion is exogenous. Our epidemiological model also complements and extends \cite{peng2020epidemic}, by allowing a diffusion dynamics similar to \cite{lloyd2006infection}. Additionally, since our network structure is not necessarily random, we are able to develop new applications. Although we only apply our model to the current COVID-19 crisis, we believe that our theory has implications for other infections that spread through physical contacts.

The remainder of this study is organized as follows: Section \ref{N-sirmodellockdown} presents the N-SIRD model with lockdown and a weighted network structure. Section \ref{infectiondynamicandlockdown} describes and solves the social planner's problem. Section \ref{sec:simulations} uses simulations to provide comparative statics analyses of our theoretical findings. Section \ref{sec:empiricalApp} provides an empirical application of the N-SIRD model with lockdown. Section \ref{sec:implicationsandconlusion} discusses some policy implications and offers concluding remarks. The Supplemental Materials contain complementary information of our N-SIRD model and additional simulation and empirical results.




\section{N-SIRD Model with the Lockdown}\label{N-sirmodellockdown}

Our Network N-SIRD model with the lockdown (or simply N-SIRD model) is an individual-based probabilistic epidemiological framework set in continuous time $t\in [0, \infty)$. We assume there is no vital dynamics so that a community of size $N$ is constant through time: $N(t) = N$ for all $t$. At any period in time $t$, individuals  are subdivided into those susceptible $S(t)$, those infected $I(t)$, those recovered $R(t)$ and those deceased $D(t)$: $S(t) + I(t) + R(t) + D(t) = N$. For simplicity, we drop the time subscript of different compartments. Each individual $i$ is in each of the four different compartments with the following probabilities: $s_i = P(i \in S)$, $x_i = P(i \in I)$, $r_i = P(i \in R)$, and $d_i = P(i\in D)$, with $s_i+ x_i + r_i + d_i =1$. Individuals move from susceptible to infected, then either recover or die. Susceptible people may become infected by coming into contact with infected individuals. We assume that physical contacts take place through an undirected weighted network, $A$. The social network structure $A$ is symmetric and it is represented by the adjacency matrix $(A_{i,j})$, where $A_{ij} = A_{ji} \in [0, \infty)$ represents the \textit{weight} or \textit{intensity} (or connection strength) at which individuals $i$ and $j$ are connected in the network $A$, with $A_{ij}= 0$ if $i=j$. One can interpret the intensity of the relationship between two individuals as the degree or frequency of interactions between these individuals. The intensity of connections is the primary source of heterogeneity between agents in the social network structure $A$. Some studies exploring virus spread in networks consider that agents with the same number of connections are similar. Then, a node in a network is a representative agent, and nodes differ by their number of connections. This literature includes a class of mean-field models and $N$-intertwined models via Markov theory in discrete and continuous times. Complementary to this literature, we consider that other characteristics may differentiate agents with the same number of connections. In Section \ref{sec:empiricalApp}, in which we apply our theory to U.S. nursing and long-term care homes \citep{chen2021nursing}, a node is a nursing home that can be either for-profit or not-for-profit, and nursing homes have different surplus functions.\footnote{One can follow \citet{Acemoglu2020}  and \citet{hashem2021matching} and extend our framework by considering other individual characteristics such as age, gender, race, health status, socio-economic status, and geographical location. This formal exercise is potential avenue for future research.}

Following the canonical SIR \citep{kermack1927contribution, kermack1932contributions} and SIRD epidemiological models\footnote{\citet{hethcote2000mathematics} and a recent textbook by \citet{brauer2012mathematical} present an overview of the class of SIRD models and some of their theoretical features in epidemiology. \citet{anastassopoulou2020data} and \citet{fernandez2020estimating} apply these models to analyze the possible outcomes of  the COVID-19 pandemic. In our framework, we assume that the contact rate $\beta$ is fixed. However, we note that other studies examine optimal non-pharmaceutical interventions to fight the COVID-19 pandemic with time-varying contact rates, including lockdown policies \citep{gollier2020pandemic}, and social distancing policies \citep{hashem2021matching, chudik2021covid}.}, we assume that agents contact other individuals in the community at a constant (passing or transmission) rate $\beta>0$. In our model, absent lockdowns, the probability of an individual $i$ being infected depends on the probability that he or she is susceptible ($s_i$), multiplied with the probability that a neighbor $j$ is infected ($x_j>0$) scaled by the connection weight ($A_{ij}>0$), and that he or she tries to infect the individual $i$ at the rate $\beta$. Infected individuals recover at rate $\gamma > 0$ or die at rate $\kappa >0$. Therefore, the infinitesimal change in infection probabilities over time of an individual $i$ in the community is:
\begin{equation*}
 \dot{x_i} =   \beta s_i \sum_{j \in N} (A_{ij} x_j) - (\gamma +\kappa) x_i.
\end{equation*}

\textbf{Lockdown}. We now incorporate a lockdown variable into the N-SIRD model to capture the fact that a social planner might decide to reduce the spread of the infection by enforcing a lockdown policy that modifies the structure of the existing social network. Let $L$ denote the lockdown state that is controlled by the social planner, and $l_i = P(i \in L)$ denote the probability that a random individual $i$ is sent into lockdown, with $l_i=1$ designating complete or full lockdown and $l_i =0$ no lockdown. Intermediate values of $l_i \in (0, 1)$ represent less extreme cases. We assume that the lockdown  policy is fully effective in curbing the contagion, i.e., complete lockdown is similar to quarantine or self-isolation. An individual in complete lockdown is disconnected from all their contacts. Thus, susceptible individuals in complete lockdown in period $t$ remain susceptible in the next period $t+\epsilon$, $\epsilon$ positive and very small. Therefore, with lockdown, the probability of an individual $i$ being infected depends on the probability that he or she is susceptible $(s_i)$ and is not sent in complete lockdown ($1- l_i >0$),  multiplied with the probability that a neighbor $j$ is not sent in complete lockdown ($1-l_j >0$) and is infected ($x_j>0$) scaled by the connection intensity ($A_{ij}>0$), and that he or she tries to infect the individual $i$ with the rate $\beta$. It follows that the infinitesimal change in infection probabilities over time of an individual $i$ in the N-SIRD model with lockdown is:\footnote{Our assumption of full effectiveness is contrary to \citet{alvarez2020simple} who consider that the lockdown is only partially effective in eliminating the transmission of the virus. \citeauthor{alvarez2020simple} justify this limitation by the fact that people can still interact in complete lockdown. We assume that being in complete lockdown sever the agent's contacts with all neighbors in the prevailing network.}
\begin{equation}\label{infection}
\dot{x_i}=\beta s_i(1 -l_i)\sum_{j\in N} A_{ij} (1-l_j)x_j - (\gamma +\kappa) x_i.
\end{equation}

\textbf{Disease Dynamics}. The equation generated by $\dot{x_i}$ describes the law of motion of the infection probabilities for individual $i$ in the community. The lockdown variable only modifies the structure of the existing network in our model. It is not a compartment variable that excludes being in other compartments. Any individual can be sent into lockdown regardless of whether the individual is susceptible, infected or recovered. An infected individual in complete lockdown can not transmit the virus through his or her social networks. For each $i\in N$, let $X_i= (x_i, s_i, r_i, d_i)^{T}$ denote agent $i$'s health characteristics in the population, where $T$ means ``transpose". We summarize the laws of motion of the variable of interests given the lockdown profile $l=(l_i)_{i\in N}$ by the following nonlinear system of ordinary differential equations:
\[
 \text{(ODE)}:
   \begin{dcases}
 \dot{x_i} = \beta s_i(1 -l_i)\sum_{j\in N} [A_{ij} (1-l_j)x_j] -(\gamma +\kappa) x_i\\
 \dot{s_i} = - \beta s_i(1 -l_i)\sum_{j\in N} [A_{ij} (1-l_j) x_j]\\
\dot{r_i} = \gamma x_i\\
\dot{d_i} = \kappa x_i\\
s_i + x_i + r_i + d_i =1
   \end{dcases}
\]
with the initial value point  $(x_i(0), s_i(0), r_i(0), d_i(0))$ such that
\begin{equation*}
x_i(0) \geq 0, \ s_i(0) \geq 0, \ r_i(0)\geq 0, \ d_i(0)\geq 0, \ \text{and} \ x_i(0) + s_i(0) + r_i(0)+ d_i(0)= 1. 
\end{equation*}

We use the N-SIRD model with lockdown (ODE) to obtain qualitative insight into the transmission dynamics of disease in a community. Before using the model to simulate disease dynamics and evaluate control strategies in the Sections \ref{sec:simulations} and \ref{sec:empiricalApp}, it is instructive to explore its basic qualitative properties. First, Proposition \ref{propdynamics} shows the existence of a solution for the system (ODE).

\begin{proposition}\label{propdynamics}
The system (ODE) admits a unique solution $\mathcal{S}^{*}= \mathcal{S}^{*}(l, A, \beta, \gamma, \kappa)$.
\end{proposition}
 
\begin{proof}
Given that $s_i=1-x_i-r_i-d_i$, for each $i\in N$, we can rewrite (ODE) as:
\[
 \text{(ODE)}:
   \begin{dcases}
 \dot{x_i} = \beta (1-x_i-r_i-d_i)(1 -l_i)\sum_{j\in N} [A_{ij} (1-l_j)x_j] -(\gamma +\kappa) x_i\\
 \dot{s_i} = - \beta (1-x_i-r_i-d_i)(1 -l_i)\sum_{j\in N} [A_{ij} (1-l_j) x_j]\\
\dot{r_i} = \gamma x_i\\
\dot{d_i} = \kappa x_i
   \end{dcases}
\]
Consider the vector-valued function $f_i (t, X_i)= (f_{i1} (t, X_i), f_{i2} (t, X_i), f_{i3} (t, X_i), f_{i4} (t, X_i))^{T}$, where
\begin{eqnarray*}
f_{i1} (t, X_i) &=& \beta (1-x_i-r_i-d_i)(1 -l_i)\sum_{j\in N} [A_{ij} (1-l_j)x_j] -(\gamma +\kappa) x_i\\
f_{i2} (t, X_i)&=& - \beta (1-x_i-r_i-d_i)(1 -l_i)\sum_{j\in N} [A_{ij} (1-l_j) x_j] \\
f_{i3} (t, X_i)&=& \gamma x_i \ \text{and}\\
f_{i4} (t, X_i)&=& \kappa x_i.
\end{eqnarray*}
The function $f_i$ is a continuously differentiable function, for each $i \in N$. Consequently, the ODE admits a unique solution, $\mathcal{S}^{*}(l, A, \beta, \gamma, \kappa, X_0)$, thanks to the theorem of existence and uniqueness of a solution for first-order general ordinary differential equations, where $l = (l_i)_{i\in N} \in [0,1]^n$ is a vector of individual lockdown probabilities.
 \end{proof}
Next, we carry our analysis of the N-SIRD model in the feasible domain: 
\begin{equation*}
    \Omega = \{( (x_i)_{i\in N}, (s_i)_{i\in N}, (r_i)_{i\in N}, (d_i)_{i\in N}) \in [0, 1]^{4n}: x_i + s_i + r_i + d_i \leq  1, 1 \leq i \leq n\}
\end{equation*}
The domain $\Omega$ is positively invariant (i.e., solutions that start in $\Omega$ remain in $\Omega$  for all $t\geq 0$). Hence, we can confirm that the system (ODE) is mathematically and epidemiologically well posed in $\Omega$ \citep{hethcote2000mathematics}.

\textbf{Equilibria and the basic reproduction number}. To find equilibria of system (ODE), we set each expression on the left-hand side of equations in (ODE) equal zero. It follows that any equilibrium point constitutes a disease free equilibrium point (DFE) in which the probability of infection is zero, i.e., $x_i =0$ for all $i\in N$. For simplicity, we analyse the disease dynamics at the DFE $E_0 = (0,...,0, 1,...,1, 0,...0,...,0)$ in a completely susceptible population. One of the most fundamental concepts in epidemiology is the basic reproduction number $R_0$, i.e., the expected number of secondary cases produced by a typical infected individual during its entire period of infectiousness in a completely susceptible population. Following \citet{diekmann1990definition} and \citet{van2002reproduction}, only the infected compartments $I$ are involved in the calculation of $R_0$, and the latter can be calculated using the method of next-generation matrix. Formally, $R_0$ is defined as the spectral radius of the next generation matrix
$FV^{-1}$, where $F$ is the matrix of the rate of generation of new infections, and $V$ is the matrix of transfer of individuals among the four health compartments. Following the notations in \citet{van2002reproduction}, from the system (ODE), we can write:
\begin{equation*}
    \dot{x_i} = \mathcal{F}_i - \mathcal{V}_i, \ \text{where}
\end{equation*}

\begin{equation*}
    \mathcal{F}_i= \beta (1-x_i-r_i-d_i)(1 -l_i)\sum_{j\in N} [A_{ij} (1-l_j)x_j], \ \text{and} \  \mathcal{V}_i= (\gamma +\kappa) x_i.
\end{equation*}
$F$ is the Jacobian matrix, and it is given by $F= [\frac{\partial \mathcal{F}_i} {\partial x_j} = \mathcal{F}_{ij}]_{E_0}$, and $V= [\frac{\partial \mathcal{V}_i} {\partial x_j} = \mathcal{V}_{ij}]_{E_0}$, where $x=(x_j)= (x_1, x_2,..., x_n)$. We have $\mathcal{F}_{ii} = -\beta (1-l_i) \sum\limits_{j\in N} [A_{ij} (1-l_j)x_j]$ and $\mathcal{F}_{ij} = \beta A_{ij} (1-x_i-r_i-d_i)(1 -l_i) (1-l_j)$ for $j\neq i$. At the equilibrium point $E_0$, it holds that $\mathcal{F}_{ii}(E_0) =0$ and $\mathcal{F}_{ij}(E_0) = \beta A_{ij} (1 -l_i) (1-l_j)$ for $j\neq i$. Since, $A_{ii}=0$, we can write 
\begin{equation*}
\mathcal{F}_{ij} (E_0) = \beta A_{ij}(1 -l_i) (1-l_j), \  \text{for} \ 1\leq i, j \leq n.    
\end{equation*} 
It is direct to have $\mathcal{V}_{ij}(E_0) = (\gamma + \kappa) \delta^{ij}$, where $\delta^{ij} = 1$ if $i=j$, and 0 otherwise. It follows that $V_{ii}= \gamma + \kappa$ and $V_{ii}^{-1}= \frac{1}{\gamma + \kappa}$, for all $1\leq i \leq n$ such that
\begin{equation*}
V= diag (V_{11},..., V_{ii},..., V_{nn}) \ \text{and} \ V^{-1}= diag (V_{11}^{-1},..., V_{ii}^{-1},..., V_{nn}^{-1}).    
\end{equation*}
Therefore, $FV^{-1} \equiv M = [M_{ij}]_{1\leq i, j \leq n}$, where $M_{ij}= \frac{\beta}{\gamma + \kappa} A_{ij} (1-l_i)(1-l_j)$, and $R_0$ which is the spectral radius of $M$ is:
\begin{equation*}
    R_0 = \rho (M):= \max \{|e|: e \ \text{is an  eigenvalue of} \ M \}.
\end{equation*}
In a fully homogeneous connected society (for example, a lattice network), it holds that $A_{ij}=1$ for all agents $i$ and $j$, and without any non-pharmaceutical intervention such as lockdown, $R_0= \frac{\beta}{\gamma + \kappa}$, the standard SIRD basic reproduction number in a model with no vital dynamics. Given that the social network structure $A$ is undirected, it holds that $A_{ij}= A_{ji}$, so that $M_{ij}= M_{ji}$, for all $i$ and $j$. Additionally, since all the values $A_{ij}$, $1-l_i$, and $1-l_j$ are real and non-negative, it follows that $FV^{-1}$ is a non-negative symmetric real matrix. Therefore,  all of its eigenvalues and eigenvectors are real. Since the diagonal of $FV^{-1}$ consists of zero, it holds that the trace of $FV^{-1}$ is zero (recall that the trace of $FV^{-1}$ is the sum of its eigenvalues). Given that the determinant of $FV^{-1}$, which is the product of its eigenvalues, is not necessarily zero, it follows that $R_0$ is positive. The following result from  \citet[Theorem 2, p. 33]{van2002reproduction} provides the asymptotic stability analysis of continuum of the disease-free equilibrium $E_0$.
\begin{proposition}\label{stability}
The continuum of $DFE$ $E_0$ of system (ODE) is locally-asymptotically stable if $R_0 <1$, but unstable if $R_0>1$. 
\end{proposition}
The proof of Proposition \ref{stability} is similar to the demonstration provided by \citet[Theorem 2, p. 33]{van2002reproduction} and is therefore omitted. Following the epidemiological literature, Proposition \ref{stability} implies that a small invasion of virus-infected agents will not generate an epidemic outbreak in society when the basic reproduction number is below 1. When $R_0 < 1$, the disease rapidly dies out if the number of infected agents in a completely susceptible population is in the region of attraction of the continuum of the disease free equilibrium $E_0$. However, when $R_0>1$, the epidemic rises to a peak and then eventually declines to zero.

\vspace{2mm}

\textbf{The final size of the pandemic}. To reflect the impact of the epidemic in a totally ``virgin" or susceptible population, we set $s_{i}(0)  \approx 1$ and assume following \citet{andreasen2011final} that $x_i(0)$ is positive with $x_i (0)  \approx 0$ for all $i\in N$. One can prove that $x_i(t) \rightarrow 0$ when $t \rightarrow \infty$, while there exists some real number $s_i(\infty)$ such that $s_i(t) \rightarrow s_i(\infty)$ when $t \rightarrow \infty$ (see, for instance, \citet{brauer2008epidemic}). The first two equations in the system (ODE) can be rewritten as:
\[
  \begin{dcases}
 \dot{x_i} = \beta s_i\sum_{j\in N} [A_{ij} (1-l_i) (1-l_j)x_j] -(\gamma +\kappa) x_i\\
 \dot{s_i} = - \beta s_i\sum_{j\in N} [A_{ij} (1-l_i) (1-l_j) x_j].\\
 \end{dcases}
\]
We may determine the value of $s_i(\infty)$ by integration of the first two equations in the system (ODE) over the entire epidemic period, which entails
\begin{equation}\label{suscepintegration}
 \text{log} \ s_i(\infty) - \text{log} \ s_i(0) = -\beta \sum_{j\in N} [A_{ij} (1-l_i) (1-l_j) \int\limits^{\infty}_{0} x_j d_t],   
\end{equation}
\begin{equation}\label{suscplusinfepintegration}
s_i(\infty) -  s_i(0) =  \int\limits^{\infty}_{0} ( \dot{x_i} +  \dot{s_i}) dt = - (\gamma + \kappa) \int\limits^{\infty}_{0} x_i d_t.
\end{equation}
We can derive the outcome of the epidemic in terms of the ratio $\sigma_i = \frac{s_i(\infty)}{s_i(0)}$, which is approximately the probability of being susceptible and remaining uninfected at the end of the epidemic, $s_i(\infty)$, given that $s_i(0) \approx 1$. We can use the column vector $\sigma = (\sigma_1,..., \sigma_n)^{T}$ to express the size of the epidemic since the infected rate for agent $i$ is $z_i = 1- \sigma_i$, and the final size of the epidemic in the whole population is $\sum\limits_{j\in N} z_j N$. For each agent $i$, the attack rate $z_i$ also equals $r_i(\infty) + d_i(\infty)$, since $x_i(t) \rightarrow 0$ when $t \rightarrow \infty$. Noting that $(1-\sigma_i) s_i(0) = s_i(0) - s_i(\infty)$, and substituting Eq.(\ref{suscplusinfepintegration}) into Eq.(\ref{suscepintegration}) yields the size of epidemic $\sigma$ as a solution of the coupled implicit equations:
\begin{equation}\label{sizepide1}
   0= \text{log} \ \sigma_i + \sum_{j\in N} [\frac{\beta}{\gamma + \kappa} A_{ij} (1-l_i) (1-l_j) s_j(0)] (1-\sigma_j) = H_i(\sigma), \ i=1,2..., n.
\end{equation}
Recall that $M_{ij}$, which is the $(i, j)$-entry in the next generation matrix is $\frac{\beta}{\gamma + \kappa} A_{ij} (1-l_i) (1-l_j)$. With $s_{i}(0) \approx 1$ for all $i$, the final size equation (\ref{sizepide1}) can be written in matrix notation with the next generation matrix $M$, the coordinate-wise log-function, and the column null vector $\overline{0}= (0, ..., 0)^{T}$ as:
\begin{equation}\label{sizepidematrix1}
\overline{0} = \text{log} \ \sigma + M (1-\sigma) = H(\sigma).
\end{equation}
Now, taking the coordinate-wise exp of equation (\ref{sizepidematrix1}) entails the alternative version of the final size equation in $z= 1- \sigma$, with $\overline{1}= (1,..., 1)^{T}$:
\begin{equation}\label{sizepidematrix2}
z = \overline{1} - \text{exp} (-M z)
\end{equation}
Following \citet{diekmann2000mathematical}, \citet{wallinga2006using}, and \citet{andreasen2011final}, we can interpret equation (\ref{sizepidematrix2}) as a probabilistic identity as $z_i$ is the probability that agent $i$ becomes infected during the epidemic while $\text{exp} (- \sum\limits_{j\in N} M_{ij} z_j)$ gives the probability of remaining susceptible during the entire epidemic. The question that remains is whether the final size equation (\ref{sizepidematrix1}) admits a solution. It is unambiguous to show that the column vector $\sigma^{0} = (1,...,1)^T$, which corresponds to the disease free equilibrium $E_0$, yields $H(\sigma^{0})=\overline{0}$, meaning that $\sigma^{0}$ is a solution to the problem (\ref{sizepidematrix1}). However, more solutions might exist to the final size equation. Using the result from \citet[Theorem 2, p. 2313]{andreasen2011final}, we provide the adapted following proposition that specifies conditions for solutions to Eq.(\ref{sizepidematrix1}).

\begin{proposition}\label{finalsizesolution}
Let $v_1,..., v_n$ denote the set of eigenvectors and generalized eigenvectors of the next generation matrix $M$ and $u_1,..., u_n$ the set of these vectors squared coordinate-wise. If each $u_k$ is linearly independent of the set of all eigenvectors and generalized eigenvectors excluding $v_k$, then the final size equation (\ref{sizepidematrix1}) has a single solution in the open unit $(0, 1)^n$ if $R_0 >1$ and none if $R_0 <1$.
\end{proposition}

The proof of Proposition \ref{finalsizesolution} follows the same reasoning as the one provided by \citet[Theorem 2, p. 2313]{andreasen2011final} and is therefore omitted. Next, we exploit the system (ODE) to provide an optimal non-pharmaceutical intervention against virus spread in the population.

\section{The Planning Problem: Optimal Lockdown} \label{infectiondynamicandlockdown}

The unique solution of the nonlinear system (ODE)  in Section \ref{N-sirmodellockdown} depends on both the social network $A$ and the lockdown variable $l$. The planning problem consists of choosing the policy instrument $l$ optimally since this enable the planner to influence the disease dynamics (or state variables). As mentioned in the Introduction, the first approach to containing the spread of COVID-19 in many countries was to enforce a quasi-complete lockdown policy. While this approach has proved to slow the spread of the virus, its economic costs have been significant. Governments around the world have been implementing less costly alternatives consisting of sending only certain individuals into lockdown while letting others go back to work. This raises the question of whether an optimal lockdown policy exists, and, if it does, whether it is unique. 

In this section, we answer this question for a social planner that \textit{prioritizes} health over short-term wealth accumulation, and we show that, under minimal conditions, there exists a unique optimally targeted lockdown policy. More formally, we assume that the social planner's problem consists of choosing the lockdown policy $l$ that: 
\begin{enumerate}
    \item  contains the infection \emph{incidence} level (or the relative number of new infections) below a \textit{tolerable} threshold  $\lambda$; and 
    \item minimizes the economic costs of the infection to the entire society, in this order of priority.
\end{enumerate}

This lexicographic objective problem is formalized below.

\textbf{Containing the spread of infection}. Using $\dot{x_i}$ in the system (ODE), the first objective of the planner is to select a lockdown policy $l$ such that:
\begin{equation}\label{inf-eq}
\dot{x_i}\equiv \dot{x_i}(l) \leq \lambda, \ \text{where} \ \lambda \ \text{is a non-negative parameter}.   \end{equation}
Note that the system (ODE) together with Eq.(\ref{inf-eq}) admits at least one solution. In fact, consider the policy $l$ where each individual is sent into complete lockdown, i.e, $l_i(t)=1$ for all $i\in N$ and $t$. Then, $\dot{x_i}(l)=-(\gamma +\kappa) x_i$. Therefore, given any $\lambda \geq 0$, it follows that $ \dot{x_i}(l) \leq \lambda$. However, this extreme solution induces significant social costs. In practice, the upper bound of the parameter $\lambda$ could be equal to the basic reproduction number without any lockdown policy, $R_0^{v} = \rho(M^v)$, where $M^v = [\frac{\beta}{\gamma + \kappa} A_{ij}]_{1\leq i, j \leq n}$. Given that lockdown implies a reduction of economic activities, an economic-focus planner might tolerate a value of $\lambda$ close to $R_0^{v}$. In contrast, a cautious (or prudent) planner who prioritizes health over short-term wealth accumulation during a pandemic may only tolerate infection incidences $\lambda$ that fall behind the basic reproduction number $R_0$.

\textbf{Minimizing the economic costs of lockdown}. The planner's second-order objective is to minimize the economic costs of lockdown by choosing from the set of policies that satisfy the first objective, i.e., the system (ODE) together with Eq.(\ref{inf-eq}), the one that maximizes the present discounted value of aggregate wealth or surplus. To assess the economic effects of lockdown in the population during a pandemic, we consider a simple production economy that we describe as follows. 

At any given period $t$, each individual $i$ possesses a capital level $k_i$, and a labor supply $h_i$. We assume, as in most SIR models, that individuals who recover from the infection are immune to the virus and must be released to the workforce. It follows that individuals in compartments $S$, $I$, and $R$ are the only potential workers in the economy. The individual labor supply depends on individuals' health compartments  and  probability of being into the lockdown: $h_i=h_i(s_i, x_i, r_i, d_i, l_i)$, with $h_i$ assumed to be continuous and differentiable in each of its input variables. We  assume that individual economic productivity is non-decreasing with being in susceptible and recovery compartments: $\frac{\partial h_i}{\partial s_i}\geq 0$, and $\frac{\partial h_i}{\partial r_i}\geq 0$. In contrast, labor supply is non-increasing with illness, death, and being in lockdown: $\frac{\partial h_i}{\partial x_i}\leq 0$, $\frac{\partial h_i}{\partial d_i}\leq 0$, and $\frac{\partial h_i}{\partial l_i}\leq 0$. Naturally, an individual who is working despite being infected and sick produces less compared to when this individual is healthy. Without loss of generality, we assume that capital is constant over time ($k_i(t)= k_i$, for each $t$), and labor is a variable input in the production. A combination of capital and labor supply generates output $y_i$ according to the following production function: $y_i= y_i(k_i, h_i) = y_i(k_i, s_i, x_i, r_i, d_i, l_i)$. We assume that $y_i$ is continuous and differentiable in each of its input variables. Moreover we make the following natural assumptions: $\frac{\partial y_i}{\partial k_i} \geq 0$, $\frac{\partial y_i}{\partial s_i} \geq 0$, $\frac{\partial y_i}{\partial x_i} \leq 0$, $\frac{\partial y_i}{\partial r_i} \geq 0$, $\frac{\partial y_i}{\partial d_i} \leq 0$, $\frac{\partial y_i}{\partial l_i} \leq 0$, and $\frac{\partial y_i^2}{\partial^2 v}\leq 0$, for each $v \in  \{k_i,  s_i, x_i, r_i,  d_i, l_i\}$. 
Other important variables of the problem include: the individual cost of one unit of labor ($w_i$), the price per unit of output ($p_i$), and  the social planner's discount rate ($\delta$). With the above information, agent $i$'s surplus function, $W_i$, is given as:
\begin{equation*}
 W_i(k_i, s_i, x_i, r_i, d_i, l_i) =  p_i y_i(k_i, s_i, x_i, r_i, d_i, l_i) - w_i h_i(s_i, x_i, r_i, d_i, l_i).   
\end{equation*}
The planner chooses the lockdown profile $l = (l_i)_{i\in N} \in [0,1]^n$ to maximize the present discounted value of aggregate surplus: 
\begin{equation*} \label{welfare}
\begin{split}
W(k, s, x, r, d, l): & = \int \limits_{0}^{\infty} e^{-\delta t} \left\{\sum \limits_{i \in N} W_i(k_i, s_i, x_i, r_i, d_i, l_i) \right\} dt\\
 & = \sum \limits_{i \in N} \left\{ \int \limits_{0}^{\infty} e^{-\delta t}\left( p_i y_i(k_i, s_i, x_i, r_i, d_i, l_i) - w_i h_i(s_i, x_i, r_i, d_i, l_i)\right) dt\right\}
\end{split}
\end{equation*}
\textbf{The social planner's problem}. Recall $X_i= (x_i, s_i, r_i, d_i)^T$ agent $i$'s health characteristics in the population. Given a tolerable infection incidence $\lambda$, the planner's task is to choose the optimal admissible lockdown path $l_i^{*}(t)$, for each agent $i\in N$, in period $t$, which along with the associated optimal admissible state path $X_i^{*}(t)$, will maximize the objective functional $W$. Using optimal control theory, we can formalize the social planner's problem as:
\begin{equation}\label{plannerproblem}
\begin{aligned}
& \underset{(l_i)_{i\in N}}{\text{Maximize}}
& & \int \limits_{0}^{\infty} e^{-\delta t} \sum \limits_{i\in N} \left\{ p_i y_i(k_i, s_i, x_i, r_i, d_i, l_i) - w_i h_i(s_i, x_i, r_i, d_i, l_i)\right\} dt \\
& \text{subject to}
& & \text{(ODE) and} \  \dot{x_i} \leq \lambda, \ i\in N \\
& \text{and} 
& & \ l_i(t) \in [0, 1] \ \text{for all} \ i \in N \ \text{and} \ t.
\end{aligned}
\end{equation}
We have the following result.
\begin{proposition}\label{prop-sp}
The social planner's problem (\ref{plannerproblem}) has a unique solution. 
\end{proposition}

\begin{proof}
We denote $f_i(k_i, s_i, x_i, r_i, d_i, l_i) \equiv f_i(k_i, x_i, r_i, d_i, l_i) = \beta [1-x_i-r_i-d_i] (1 -l_i) \sum\limits_{j \in N} [A_{ij} (1-l_j) x_j] -(\gamma +\kappa) x_i$, and $W_i(k_i, s_i, x_i, r_i, d_i, l_i)= p_i y_i(k_i, s_i, x_i, r_i, d_i, l_i) - w_i h_i(s_i, x_i, r_i, d_i, l_i)$. The function $l_i: t \longrightarrow l_i(t) \in [0,1]$ is continuous. The function $W_i$, and the objective function in  (\ref{plannerproblem}) are continuous and differentiable. Moreover, $f_i$ and the right-hand sides of the laws of motion in  (\ref{plannerproblem}) are all continuous and differentiable. It follows that the problem (\ref{plannerproblem}) admits a unique  optimal path $\{l^{*}(t)\}$ of the control variable (and the states $\{x^{*}(t), r^{*}(t), d^{*}(t), s^{*}(t) \}$, given the initial conditions $X_0$ and the laws of motion).
\end{proof}

Proposition \ref{prop-sp} states the existence and uniqueness of a solution to the social planner's problem. In what follows, we extend the analysis of problem (\ref{plannerproblem}) that proves useful in showing how we obtain our simulated results.  The current Hamiltonian of problem (\ref{plannerproblem}) is given as:
 \begin{multline*} 
     \mathcal{H}_c (l, x, r, d, s, \mu^1,\mu^2,\mu^3, \mu^4)= \sum \limits_{i\in N} W_i(k_i, s_i, x_i, r_i, d_i, l_i) + \sum \limits_{i\in N} \mu_i^{1} f_i + \gamma \sum  \limits_{i\in N} \mu_i^{2} x_i  + \kappa \sum \limits_{i\in N} \mu_i^{3} x_i  \\ \nonumber
    + \sum \limits_{i\in N} \mu_i^{4} [-f_i-(\gamma + \kappa) x_i],
 \end{multline*}
where $\mu_i^j$ ($j=1,..,6$), for each $i\in N$, are the costate variables. Given the inequality constraints $\dot{x_i} \leq \lambda$, and the constraints $l_i(t) \in [0, 1]$, we can augment the current Hamiltonian $\mathcal{H}_c$ into the current Lagrangian function:
\begin{multline*} 
     \mathcal{L}_c(l, x, r, d, s, \mu^1,\mu^2,\mu^3, \mu^4, \theta^1, \theta^2, \theta^3)= \sum \limits_{i\in N} W_i(k_i, s_i, x_i, r_i, d_i, l_i) + \sum \limits_{i\in N} \mu_i^{1} f_i + \gamma \sum  \limits_{i\in N} \mu_i^{2} x_i  + \kappa \sum \limits_{i\in N} \mu_i^{3} x_i \\ \nonumber
    + \sum \limits_{i\in N} \mu_i^{4} [-f_i-(\gamma + \kappa) x_i] + \sum \limits_{i\in N} \theta_i^{1} (\lambda - f_i) + \sum \limits_{i\in N} \theta_i^{2}l_i + \sum \limits_{i\in N} \theta_i^{3} (1-l_i), 
 \end{multline*}
where the parameters $\theta^j$, $j=1,2,3$, are Lagrange multipliers. We can also rewrite $\mathcal{L}_c$ as:
  \begin{multline*} 
     \mathcal{L}_c(l, x, r, d, s,  \mu^1,\mu^2,\mu^3, \mu^4, \theta^1, \theta^2, \theta^3)= \sum \limits_{i\in N} W_i(k_i, s_i, x_i, r_i, d_i, l_i) + \sum \limits_{i\in N} (\mu_i^{1}-\mu_i^{4}-\theta_i^{1}) f_i  + \gamma \sum  \limits_{i\in N} \mu_i^{2} x_i \\ \nonumber
 + \kappa \sum \limits_{i\in N} \mu_i^{3} x_i -(\gamma+\kappa) \sum \limits_{i\in N} \mu_i^{4} x_i + \lambda \sum \limits_{i\in N} \theta^1_{i} + \sum \limits_{i\in N} \theta_i^{2}l_i + \sum \limits_{i\in N} \theta_i^{3} (1-l_i). 
 \end{multline*}
 The first-order conditions for maximizing  $\mathcal{L}_c$ call for, assuming interior solutions,
\begin{equation} \label{foclagrange1}
     \frac{\partial \mathcal{L}_c}{\partial l_k}= 0, \; k \in N, 
 \end{equation}
as well as for each $k\in N$:

\begin{align}
 \frac{\partial \mathcal{L}_c}{\partial \theta^{1}_k}&= \lambda - \dot{x_k}\geq 0,          &  \theta^{1}_k& \geq 0,       &  \theta^{1}_k \frac{\partial \mathcal{L}_c}{\partial \theta^{1}_k}&= \theta^{1}_k (\lambda - \dot{x_k})=0, \label{foclagrangemu1}\\
\frac{\partial \mathcal{L}_c}{\partial \theta^{2}_k}&= l_k\geq 0,          &  \theta^{2}_k& \geq 0,       &  \theta^{2}_k \frac{\partial \mathcal{L}_c}{\partial \theta^{2}_k}&= \theta^{2}_k l_k=0, \ \text{and}\label{foclagrangemu2}\\
\frac{\partial \mathcal{L}_c}{\partial \theta^{3}_k}&= 1 - l_k\geq 0,          &  \theta^{3}_k& \geq 0,       &  \theta^{3}_k \frac{\partial \mathcal{L}_c}{\partial \theta^{3}_k}&= \theta^{3}_k (1 - l_k)=0.\label{foclagrangemu3}
\end{align}

Finally, the other maximum-principle conditions that include the dynamics  for state and co-state variables are, for $\; k \in N$: 
\begin{align*}
\dot{x_k}&= \frac{\partial \mathcal{L}_c}{\partial \mu_k^{1}}           &  \dot{r_k}&= \frac{\partial \mathcal{L}_c}{\partial \mu_k^{2}}             &  \dot{d_k}&= \frac{\partial \mathcal{L}}{\partial \mu_k^{3}}  &
\dot{s_k}&= \frac{\partial \mathcal{L}_c}{\partial \mu_k^{4}}, \ \text{and}
\end{align*}
\begin{align}
\dot{\mu_k^{1}}&= \delta \mu_k^{1}-\frac{\partial \mathcal{L}_c}{\partial x_k}   & \dot{\mu_k^{2}}&=\delta \mu_k^{2}-\frac{\partial \mathcal{L}_c}{\partial r_k}          &  \dot{\mu_k^{3}}&=\delta \mu_k^{3}-\frac{\partial \mathcal{L}_c}{\partial d_k}  &  \dot{\mu_k^{4}}&=\delta \mu_k^{4}-\frac{\partial \mathcal{L}_c}{\partial s_k} \label{foclagrange4}
\end{align}

Recall that $f_i(x_i, r_i, d_i, l_i) = \beta (1-x_i-r_i-d_i) (1 -l_i) \sum\limits_{j \neq i} [A_{ij} (1-l_j) x_j] -(\gamma +\kappa) x_i$. Then,
\begin{align*} 
 \frac{\partial f_i}{\partial l_k} &= 
 \begin{cases} 
   -\beta (1-x_i-r_i-d_i) \sum\limits_{j \neq i} [A_{ij} (1-l_j) x_j] & \text{if } k = i \\
   -\beta (1-x_i-r_i-d_i) (1 -l_i) A_{ik}x_k & \text{if } k \neq i
  \end{cases}  \\ 
\frac{\partial f_i}{\partial x_k} &=  \begin{cases} 
  -\beta (1-l_i)\sum\limits_{j \neq i} [A_{ij} (1-l_j) x_j] -(\gamma+\kappa) & \text{if } k = i \\
\beta (1-x_i-r_i-d_i) (1 -l_i)(1-l_k) A_{ik}    & \text{if } k \neq i
\end{cases} 
 \frac{\partial f_i}{\partial r_k} =  \frac{\partial f_i}{\partial d_k} &= \begin{cases} 
  -\beta (1-l_i)\sum\limits_{j \neq i} [A_{ij} (1-l_j) x_j] & \text{if } k = i \\
   0 & \text{if } k \neq i
  \end{cases} 
\end{align*}

We also recall that $W_i(k_i, s_i, x_i, r_i, d_i, l_i)=p_i y_i(k_i, s_i, x_i, r_i, d_i, l_i) - w_i h_i(s_i, x_i, r_i, d_i, l_i)$. Therefore, for each $i$ and $k$, and for each $u\in \{s_k, x_k, r_k,  d_k, l_k\}$, it holds that
\begin{align} 
 \frac{\partial W_i}{\partial u} &= 
 \begin{cases} 
   p_i  \frac{\partial y_i}{\partial u} - w_i \frac{\partial h_i}{\partial u} & \text{if } k = i \\
   0 & \text{if } k \neq i
  \end{cases} \label{derWtlxd}
\end{align}

Therefore, for each $k\in N$, we can write $\frac{\partial \mathcal{L}_c}{\partial l_k}$ as:
\begin{equation} \label{derLwrtlk}
\begin{split}
\frac{\partial \mathcal{L}_c}{\partial l_k} & = \sum\limits_{i \in N} \frac{\partial W_i}{\partial l_k} + \sum\limits_{i \in N}(\mu^1_{i}-\mu^4_{i}-\theta_i^{1})\frac{\partial f_i}{\partial l_k} - \theta_k^{2} - \theta_k^{3} \\
 & = \frac{\partial W_k}{\partial l_k} + \sum\limits_{i \in N}(\mu^1_{i}-\mu^4_{i}-\theta_i^{1})\frac{\partial f_i}{\partial l_k} - \theta_k^{2} - \theta_k^{3} \ \text{using} \ (\ref{derWtlxd})\\
 & =  p_k  \frac{\partial y_k}{\partial l_k} - w_k \frac{\partial h_k}{\partial l_k} + \sum\limits_{i \in N}(\mu^1_{i}-\mu^4_{i}-\theta_i^{1})\frac{\partial f_i}{\partial l_k} - \theta_k^{2} - \theta_k^{3}
\end{split}
\end{equation}
 Hence, using the first-order conditions (\ref{foclagrange1}), equation (\ref{derLwrtlk}) becomes:
 \begin{equation*}\label{foc}
 0=p_k  \frac{\partial y_k}{\partial l_k} - w_k \frac{\partial h_k}{\partial l_k} + \sum\limits_{i \in N}(\mu^1_{i}-\mu^4_{i}-\theta_i^{1})\frac{\partial f_i}{\partial l_k} - \theta_k^{2} - \theta_k^{3}.    
 \end{equation*}

Using the other conditions from (\ref{foclagrange4}) and using (\ref{derWtlxd}): 
\begin{equation} \label{costate1}
\dot{\mu_k^{1}} = \delta \mu_k^{1}-\frac{\partial \mathcal{L}_c}{\partial x_k} =  \delta \mu_k^{1}-p\frac{\partial y_k}{\partial x_k} + w_k \frac{\partial h_k}{\partial x_k} - \gamma \mu_k^{2} - \kappa \mu_k^{3}+(\gamma+\kappa) \mu^4_{k} -\sum\limits_{i\in N} (\mu^1_{i}-\mu^4_{i}-\theta_i^{1}) \frac{\partial f_i}{\partial x_k}.
\end{equation}

Similarly, using (\ref{foclagrange4}), we get:
\begin{equation} \label{costate2}
\dot{\mu_k^{2}}=\delta \mu_k^{2}-\frac{\partial \mathcal{L}_c}{\partial r_k} =\delta \mu_k^{2} -p\frac{\partial y_k}{\partial r_k} + w_k \frac{\partial h_k}{\partial r_k}-\sum\limits_{i\in N} (\mu^1_{i}-\mu^4_{i}-\theta_i^{1})\frac{\partial f_i}{\partial r_k} \ \text{using} \ (\ref{derWtlxd}), 
\end{equation}

\begin{equation}\label{costate3} 
\dot{\mu_k^{3}}=\delta \mu_k^{3}-\frac{\partial \mathcal{L}_c}{\partial d_k} =\delta \mu_k^{3} -p\frac{\partial y_k}{\partial d_k} + w_k \frac{\partial h_k}{\partial d_k} -\sum\limits_{i\in N} (\mu^1_{i}-\mu^4_{i}-\theta_i^{1})\frac{\partial f_i}{\partial d_k} \ \text{using} \ (\ref{derWtlxd}),     
\end{equation}
and
\begin{equation}\label{costate4}
\dot{\mu_k^{4}}=\delta \mu_k^{4}-\frac{\partial \mathcal{L}_c}{\partial s_k}=\delta \mu_k^{4}-p\frac{\partial y_k}{\partial s_k} + w_k \frac{\partial h_k}{\partial s_k}. 
\end{equation}
Note that determining a closed-form solution of the planning problem (\ref{plannerproblem}) is intractable. This is justified by the complexity and the stochastic nature of the system (ODE) that characterizes our N-SIRD model. To gain some insight into the optimal lockdown policy and the resulting disease and costs dynamics , we follow \citet{alvarez2020simple}, \citet{Acemoglu2020}, and \citet{gollier2020cost}, and resort to simulations in Section \ref{sec:simulations}. First, in Section \ref{tradeoff}, we vary the tolerable infection incidence $\lambda$ to illustrate the tradeoff between health and wealth. Contrary to \citet{bosi2021optimal} who propose a constant optimal lockdown policy to curve the contagion, our lockdown is dynamic, and more in line with \citet{alvarez2020simple}, \citet{Acemoglu2020}, and \citet{gollier2020cost}. We differ from \citet{alvarez2020simple} and \citet{Acemoglu2020} by not constraining the lockdown probability by an upper bound less than one, which situates our study more in line with \citet{bosi2021optimal}. A planner could lock down all the society if they found it optimal. Though this case corresponds to a pure epidemiological model, our findings illustrate that complete lockdown is not an optimal solution. Second, in Section \ref{illustration1}, by changing the nature of the network structure $A$, we illustrate how network configuration affects the disease dynamics and their impact on the economy. Similarly, we also illustrate in Section \ref{centralitylockdown} the effects of network centrality on individual lockdown probabilities.

\section{Comparative Statics: A Simulation Analysis}\label{sec:simulations}

We choose the N-SIRD model's parameters to match the dynamics of the infection and early studies on the COVID-19 pandemic and the period in which the researchers at the Protect Nursing Homes gathered the data on U.S. nursing homes. Following \cite{alvarez2020simple}, we use data from the World Health Organization (WHO) made public through the Johns Hopkins University Center for Systems Science and Engineering (JHU CCSE). The parameter $\beta$, the probability that an infected individual transmits the virus to a susceptible individual in their network is assumed to be 0.2. The lifetime duration of the virus is assumed to be 18 days (see, for example, \cite{Acemoglu2020} and the references therein). Given the information from JHU CCSE access on May 5, 2020, the proportion of recovered closed cases was  around  70\% for  the  U.S.,  93\%  for  Germany,  and 86\% for Spain. Thus, we assume that the parameter governing the recovery of an infected patient is given by $\gamma = \frac{0.8}{18}$, and the parameter governing the death dynamics is given by $\kappa = \frac{0.2}{18}$. We also consider the following functional form for the labour function ($h$) and the production function ($Y$):
\begin{align}
h_i(s_i, x_i, r_i, d_i, l_i) & = (1 + \phi_i s_i r_i (1-x_i)(1-d_i)) (1- \varphi_i l_i), \ (\phi_i, \varphi_i) \in [0,1]^2 \label{labfunct1}\\
y_i(k_i, s_i, x_i, r_i, d_i, l_i) & = k_i^{\alpha_i}h_i^{1-\alpha_i}, \label{prodfunct1}
\end{align}
where $\phi_i$ determines the direct effect in the rate of change in labor supply when individual $i$ is in one of the natural health compartments, $S$, $I$, $R$ or $D$, and $\varphi_i$ represents the direct effect in the decrease in labor supply when individual $i$ is in lockdown. When $d_i = P(i \in D)= 1$, we should have $l_i=0$ so that $h_i (s_i, x_i, r_i, 1, 0)=0$. In Eq. (\ref{prodfunct1}), $\alpha_i$ is the elasticity of output with respect to the capital, and $1-\alpha_i$ is the elasticity of output with respect to labor. The functions $h_i$ in Eq. (\ref{labfunct1}) and $y_i$ in Eq. (\ref{prodfunct1}) satisfy the standard conditions that we mention in Section \ref{infectiondynamicandlockdown}. 

Our choice of the Cobb-Douglas function as a parametric estimate of the production function is motivated by our empirical analyses in Section \ref{sec:empiricalApp}. Our consideration is also more in line with several studies that argue that the Cobb-Douglas function is a standard parameterization of production function in the literature \citep{douglas1976cobb}, and especially in primary care \citep{wichmann2020nonparametric}, and nursing homes \citep{reyes2020using}. Using the recent data collected by \cite{chen2021nursing} on U.S. nursing homes, we approximate a typical nursing home's production function as $y_i = k_i^{\alpha_i} h_i^{1-\alpha_i}$, where $y_i$ is the total number of residents (proxies the nursing home's output) who receive care, $k_i$ is the total number of beds (proxies the capital), and $h_i$ is the number of occupied beds (proxies the labor supply).\footnote{For more details on our estimation approach of a nursing home's production function, we refer to our Supplemental Material.} 

In all the simulations, we consider $\phi_i = 0$, and $\varphi_i=1$ such that $h_i(s_i, x_i, r_i, d_i, l_i) \approx (1-l_i)$ and we have a stationary working population. In the context of nursing and long-term care homes, we can justify the labor supply's approximation, $h_i = 1-l_i$. The connection between two nursing homes is determined by at least one signal received from a smartphone in both houses. Given the structure of U.S. nursing homes staffing practices and regulation as documented by \citet{chen2021nursing}, most of the workers in nursing homes would not be able to work remotely if the nursing home is in complete lockdown, i.e., $h_i = 0$ when $l_i =1$. Then, the choice $h_i = 1-l_i$ is a good candidate since it satisfies all the standard conditions mentioned above and allows a smooth  computational time process during our simulation analyses. As for the surplus function, we assume that $\alpha_i= \frac{1}{3}$, $p_i = 1.2$, $w_i=\frac{1}{3} p_i$, for each agent $i$,  and the level of capital is the same for all agents at all time period and normalized to $k_i=1$. The annual interest rate is assumed to be equal to 5\%. For simplicity, in our simulation analyses, we assume that networks are represented by binary adjacent matrices $A=(A_{ij})$, where $A_{ij}=1$ if agents $i$ and $j$ are connected, and $A_{ij}=0$, otherwise. Though we choose the values of the parameters of the production function to match as possible the data from the U.S. nursing and long-term care homes \citep{chen2021nursing}, we select other parameters in the planning problem for illustration purposes. Thus, one should interpret the quantitative outcomes of our N-SIRD model with caution. Nevertheless, in the empirical application, we choose the economic parameters to match each U.S. state's reality as possible.

\subsection{Infection Incidence Control and Optimal Lockdown Policy---The Health-vs-Wealth Tradeoff}\label{tradeoff}
 
\begin{figure}[H] 
\begin{subfigure}{0.5\textwidth}
\centering
\includegraphics[width=1\linewidth]{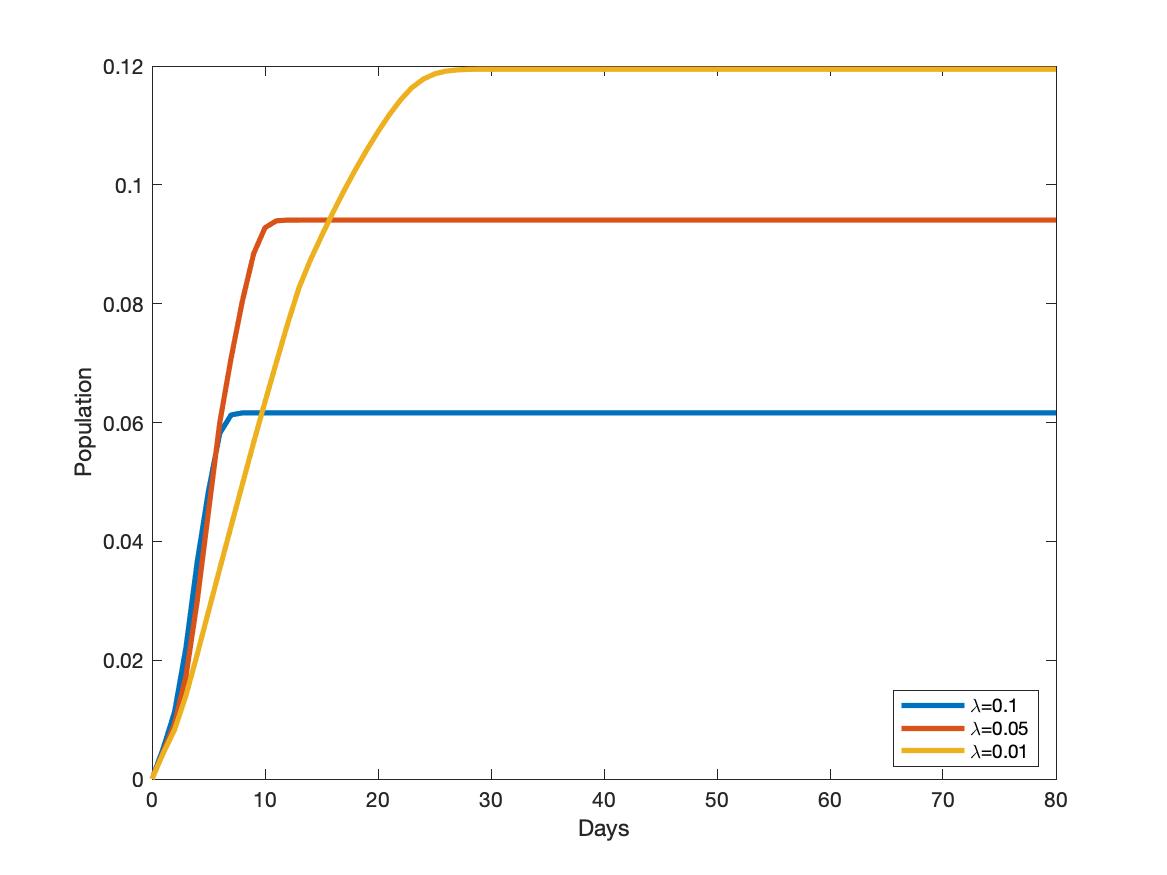}
\caption{Dynamics of Lockdown}\label{incidenceandlockdown}
\end{subfigure}
\begin{subfigure}{0.5\textwidth}
\centering
\includegraphics[width=1\linewidth]{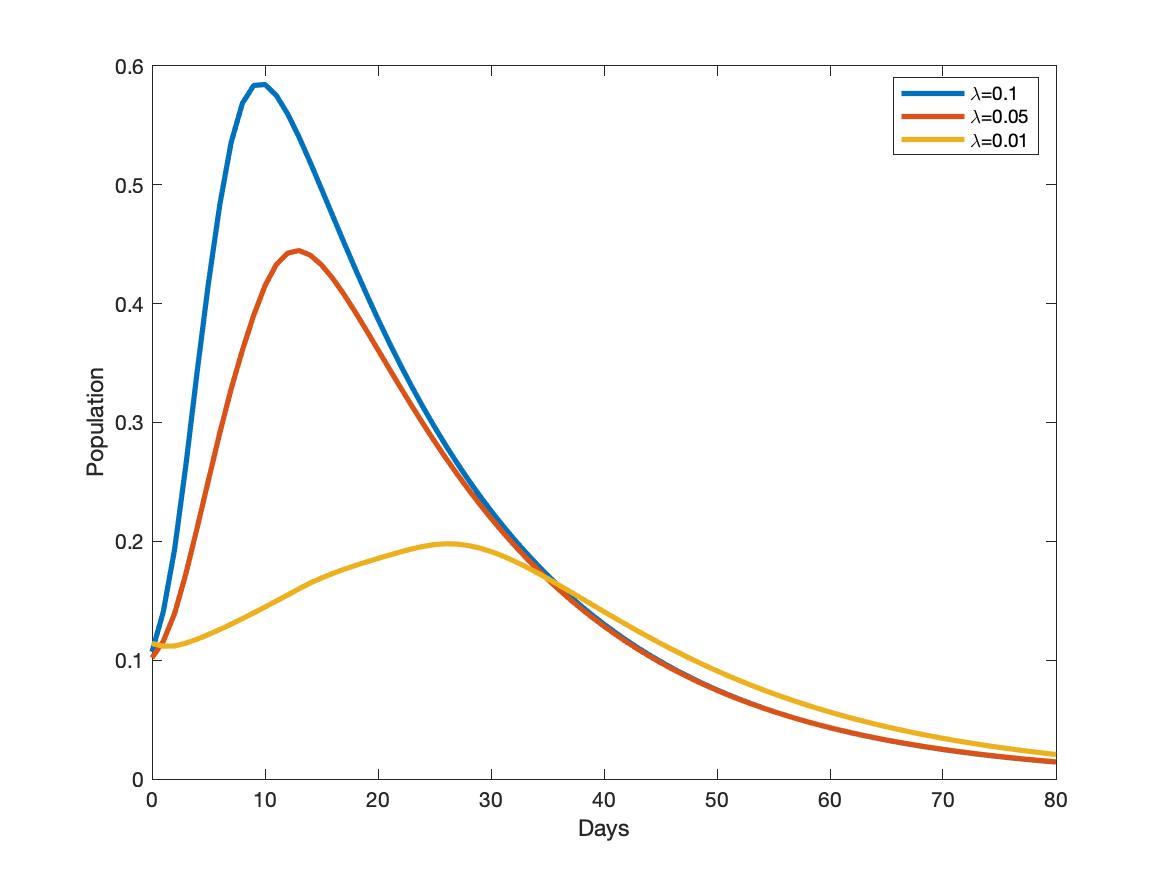}
\caption{Dynamics of Infection}\label{incidenceandinfection}
\end{subfigure}
\begin{subfigure}{0.5\textwidth}
\centering
\includegraphics[width=1\linewidth]{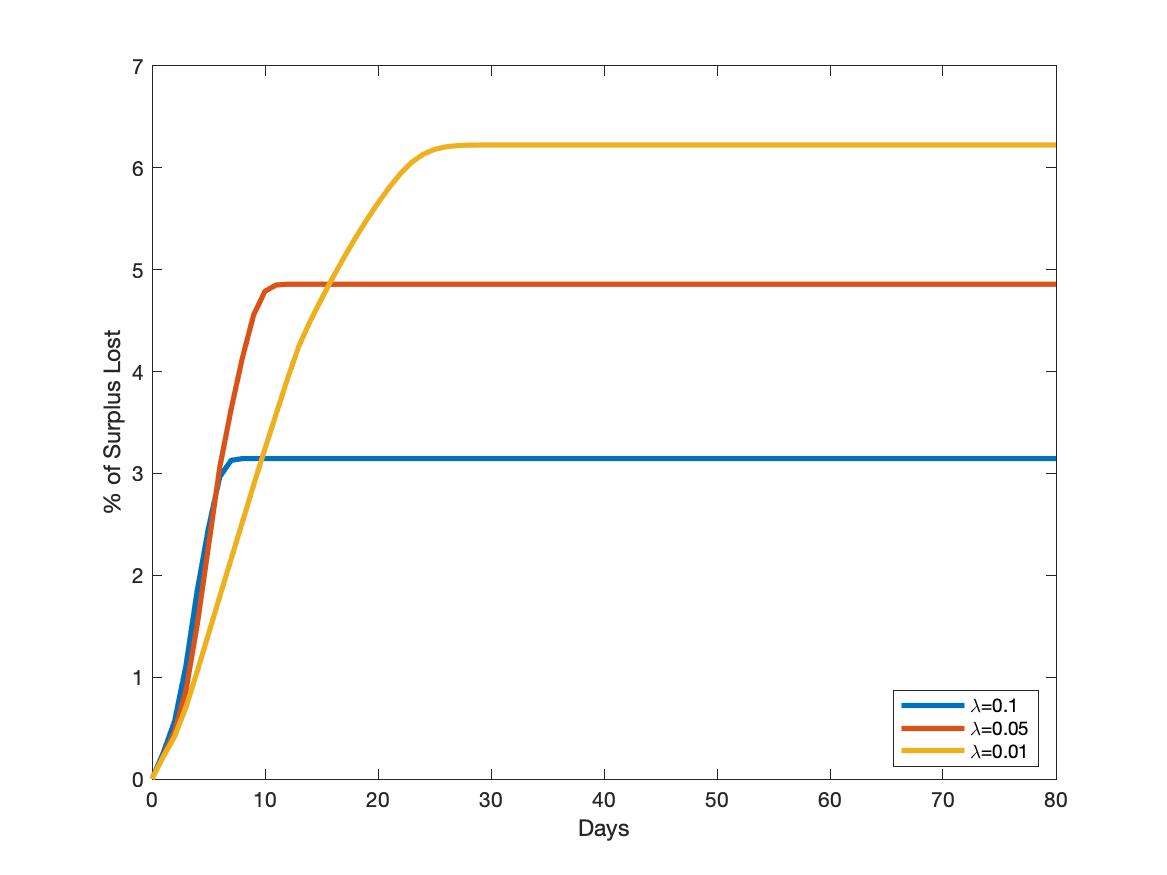}
\caption{ Dynamics of Economic Cost}\label{incidenceandcosts}
\end{subfigure}
\begin{subfigure}{0.5\textwidth}
\centering
\includegraphics[width=1\linewidth]{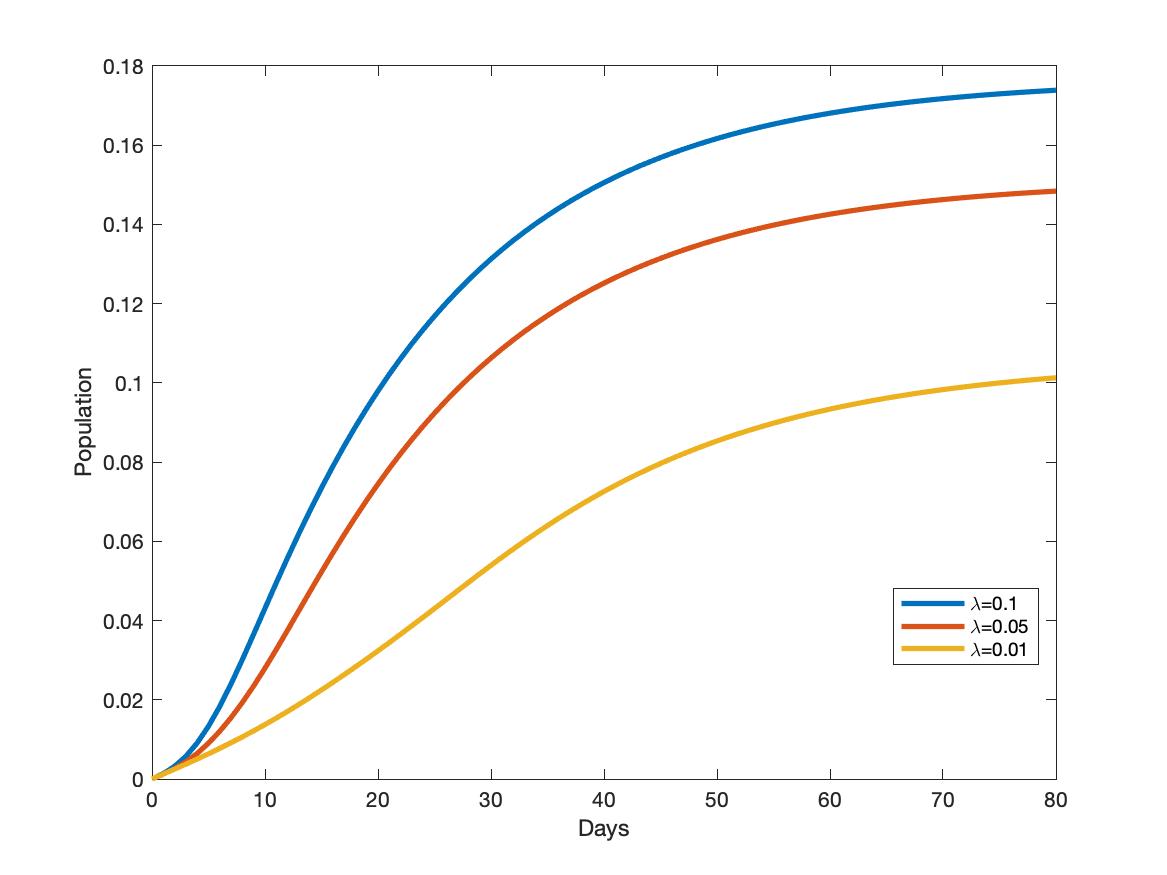}
\caption{ Dynamics of Death}\label{incidenceanddeaths}
\end{subfigure}
\caption{\textbf{Health versus Wealth tradeoff in a small-world network}. \small{We perform three sets of simulations with three different values of the tolerable infection incidence $\lambda$: 0.01, 0.05, and 0.1. The results are displayed in a two-dimensional graphic, with days in the horizontal axis, and the percentage of population affected for the variable (infection, lockdown, or death) illustrated on the vertical axis. In each period, a point in the graphic represents the average value of individual probabilities. For the economic cost, the vertical axis represents the percentage of economic (or surplus) lost relative to the economy without  the pandemic. Each graph shows three curves corresponding to three dynamics for a single variable of interest for a given value of $\lambda$. All variability within each curve in each graph is a result of the stochastic nature of transmission and not variation in the network nor $\lambda$. In the Supplemental Materials, we replicate the simulation results in Figure \ref{lock_dyn_sw} for scale-free, random, and lattice networks, in Figures 1, 2, and 3, respectively. We also replicate Figure \ref{lock_dyn_sw} using recent epidemiological data of the COVID-19 Delta variant (see Figure 4 in the Supplement Materials). We find that all these additional simulation results are qualitatively consistent with the lockdown, disease, and economic costs dynamics in Figure  \ref{lock_dyn_sw}.}}\label{lock_dyn_sw}
\end{figure}

In our first comparative statics analysis, we illustrate the effect of varying the tolerable infection incidence level on the optimal lockdown policy and describe the tradeoff between the desired level of population health and short-term economic gains. To achieve that goal, we consider an economy of $n=1000$ agents connected through a \textit{small-world network} \citep{watts1998collective} with  $2\times n$ edges ($A$ is fixed), and we vary the tolerable infection incidence, $\lambda$: 0.01, 0.05, and 0.1, in the planning problem. Figure \ref{lock_dyn_sw} represents the simulation results. 

Figure \ref{incidenceandlockdown} illustrates that the optimal cumulative lockdown rate increases with lower infection incidence level. This rate varies from around 6 percent for an incidence level equal to 0.1 to 9 percent for an incidence level of 0.05 to 12 percent for an incidence level of 0.01. What emerges from these numbers is that the relationship between the tolerable incidence level and the ultimate proportion of the population sent into lockdown is not linear. As the tolerable infection incidence level decreases, the fraction of the population sent into lockdown increases in a proportion that is lower than the decrease. The optimal lockdown policy resulting from a given tolerable infection incidence level translates into a corresponding dynamics of infection, death and economic cost. In particular, Figure \ref{incidenceandinfection} shows that a lower tolerable incidence level results in a lower infection and death rate  (see Figure \ref{incidenceandinfection} and Figure \ref{incidenceanddeaths}). Figure \ref{incidenceandcosts} illustrates the tradeoff between population health and the well-being of the economy. A lower tolerable infection incidence level induces a greater economic cost of the pandemic. Indeed, if the tolerable infection  incidence level is low, more individuals must be sent into lockdown. Then, with a decrease in individuals' productiveness in the economy, the loss in terms of economic surplus is significant. For instance, when the tolerable incidence decreases from 0.1 to 0.05, the fraction of the economic surplus lost to the pandemic increases from around 3 percent to over 5 percent; and a further decrease of the tolerable incidence level to 0.01 induces an economic surplus loss of around 6 percent. It follows that a greater level of health is achieved at the expense of short-term economic gains.     
 
\subsection{The Role of Network Configuration}\label{illustration1} 

\begin{figure}[H] 
\begin{subfigure}{0.5\textwidth}
\centering
\includegraphics[width=1\linewidth]{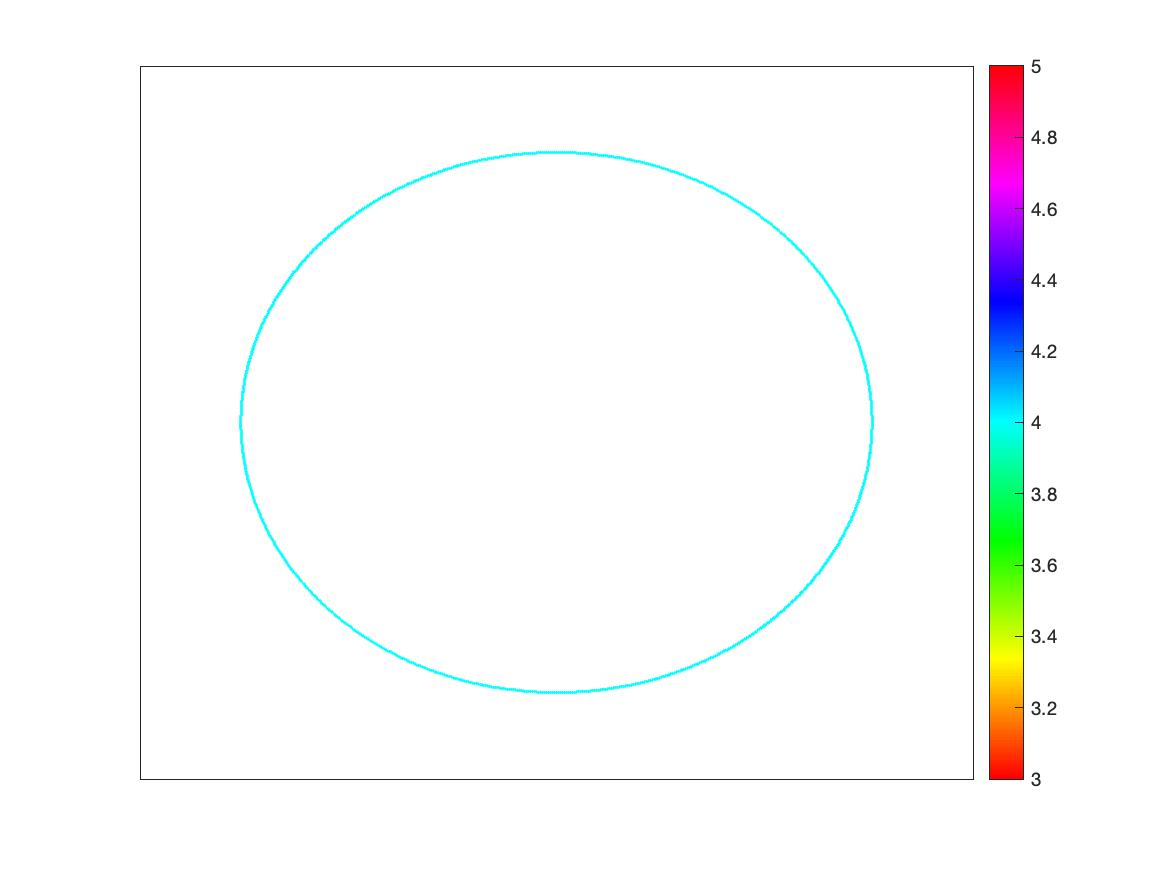}
\caption{Lattices Network }\label{latnet}
\end{subfigure}
\begin{subfigure}{0.5\textwidth}
\centering
\includegraphics[width=1\linewidth]{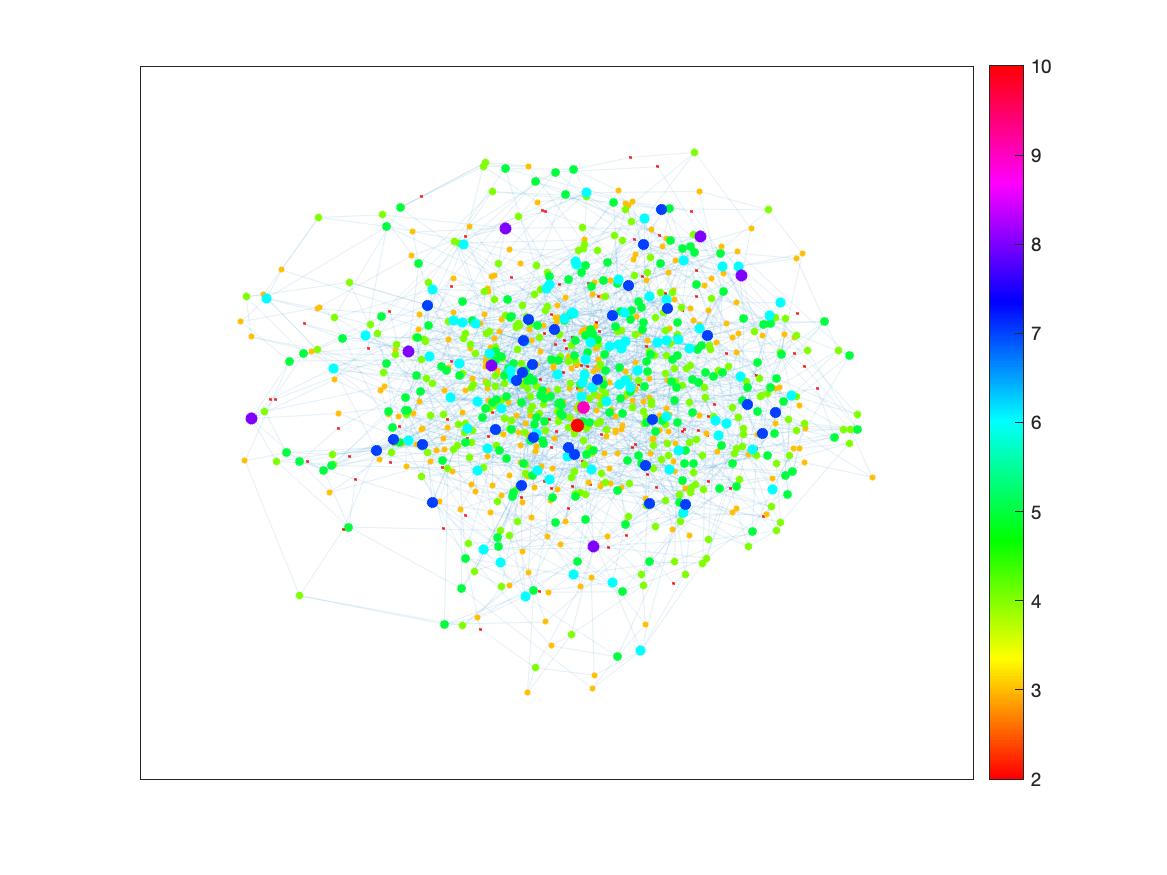}
\caption{Small-World Network}\label{smallwnet}
\end{subfigure}
\begin{subfigure}{0.5\textwidth}
\centering
\includegraphics[width=1\linewidth]{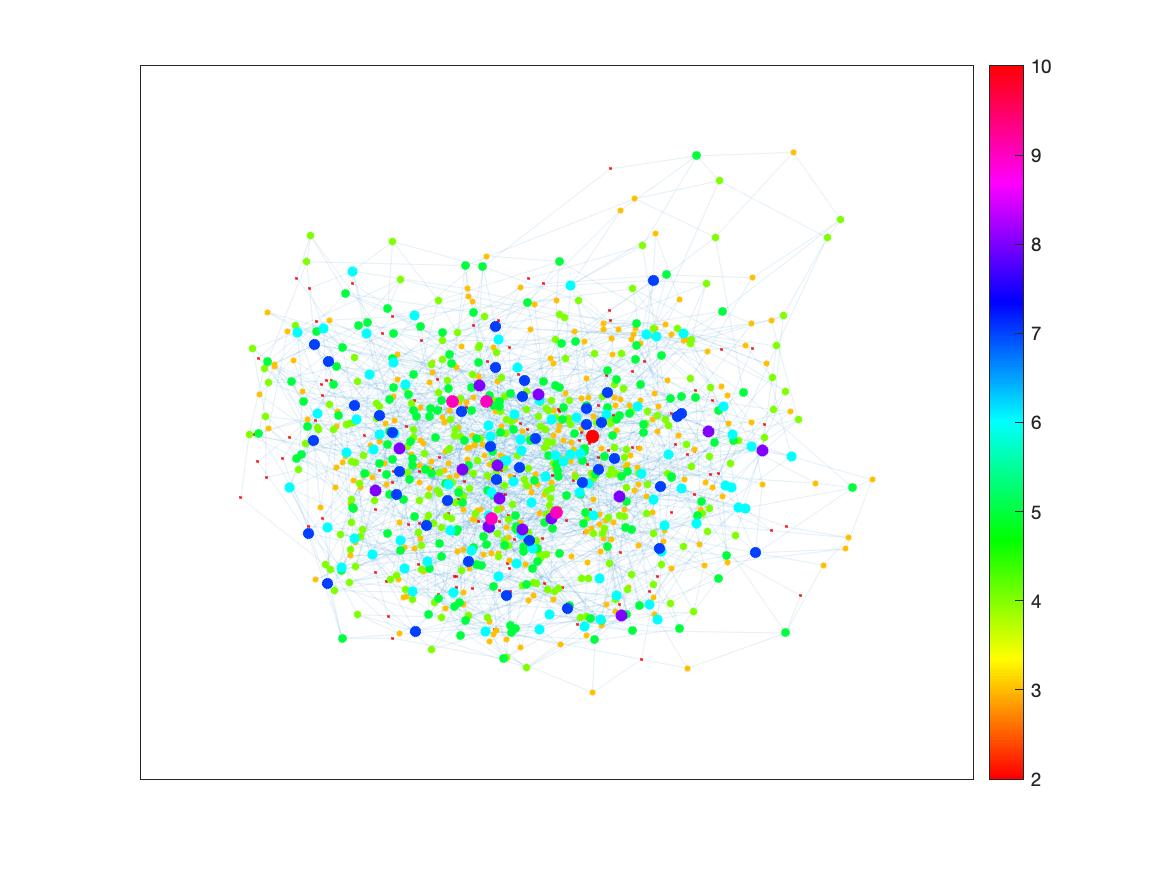}
\caption{Random Network}\label{randomnet}
\end{subfigure}
\begin{subfigure}{0.5\textwidth}
\centering
\includegraphics[width=1\linewidth]{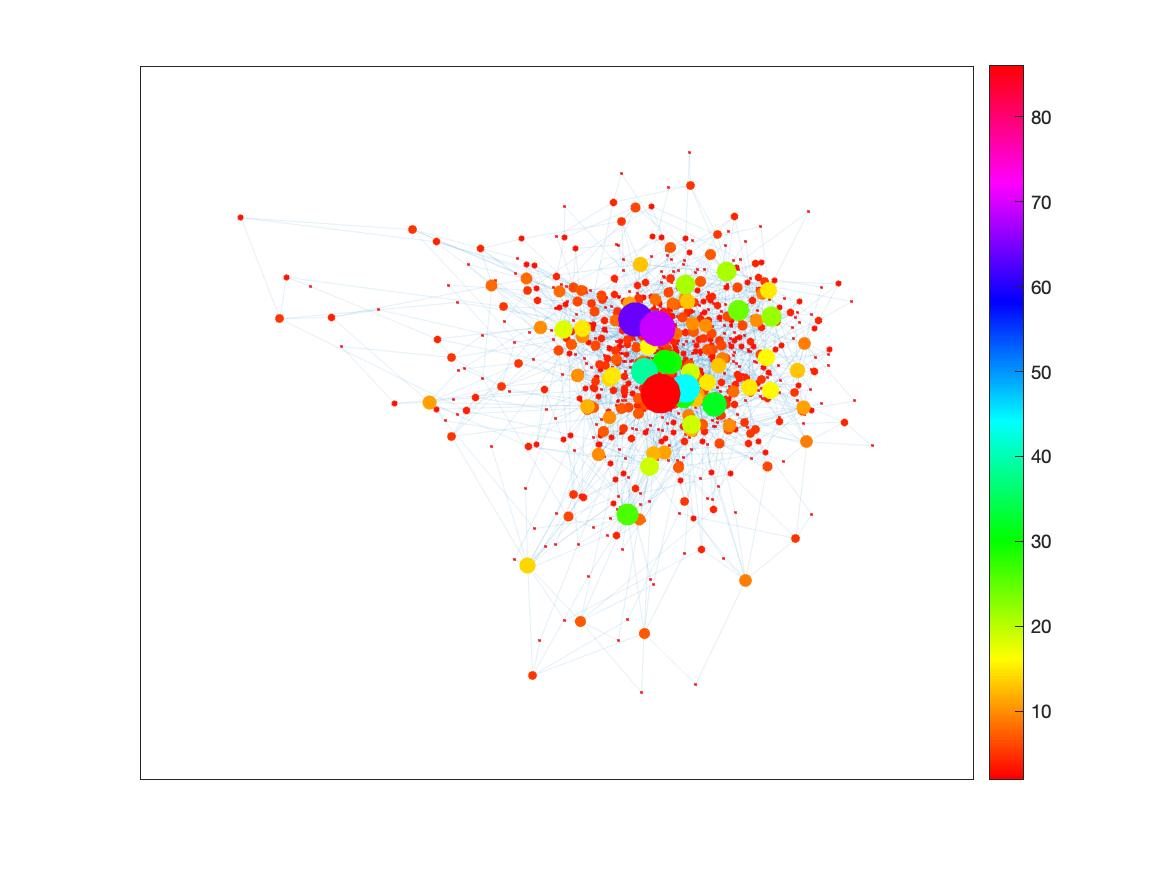}
\caption{Scale-Free Network}\label{scalefnet}
\end{subfigure}
\caption{\textbf{Simple network structures}. \small{Four distinct network types containing 1000 agents. Random networks display homogeneity of agent-level network properties and low clustering. Lattices are homogeneous at the agent level, and they show high clustering. Lattice networks also exhibit long path lengths, i.e., it takes many steps to move between two randomly selected agents, whereas random networks have short path lengths. Small-world networks display high clustering and short path lengths. Scale-free networks capture different levels of heterogeneity (for example, super-spreaders) in populations. In all four graphs, the average number of contacts per agent is 2. In each network, we represent agents with high contacts by larger dots, and we shade each node according to its number of direct contacts using the scale beside each graph.}}\label{net_struc}
\end{figure}

In Section \ref{illustration1}, we fix the tolerable infection incidence $\lambda$ to 0.01, and we vary the structure of network configuration, $A$, in the planning problem. For the sake of concreteness, we contrast four \textit{idealized} network configurations \citep{keeling2005networks}, namely a \textit{lattice network} (Figure \ref{latnet}),  a \textit{small-world network} (Figure \ref{smallwnet}), a  \textit{random network} (Figure \ref{randomnet}) and a \textit{scale-free network} (Figure \ref{scalefnet}). These networks belong to the range of the most popular network types studied in the context of disease transmission (see, for example, \citet{keeling2005networks} and the references therein for a review of these networks). According to \citeauthor{keeling2005networks}, ``Each of these idealized networks can be defined in terms of how individuals are distributed in space (which may be geographical or social) and how connections are formed, thereby simplifying and making explicit the many and complex processes involved in network formation within real populations" \citep[p. 299]{keeling2005networks}. Following this viewpoint, the networks in Figure \ref{net_struc} represent four societies of 1000 agents each that are identical in all ways except the configuration of their contact network. The four network configurations differ in both clustering of connections and path lengths between nodes, two essential factors in disease spread.

\begin{figure}[H]
\begin{subfigure}{0.5\textwidth}
\centering
\includegraphics[width=1\linewidth]{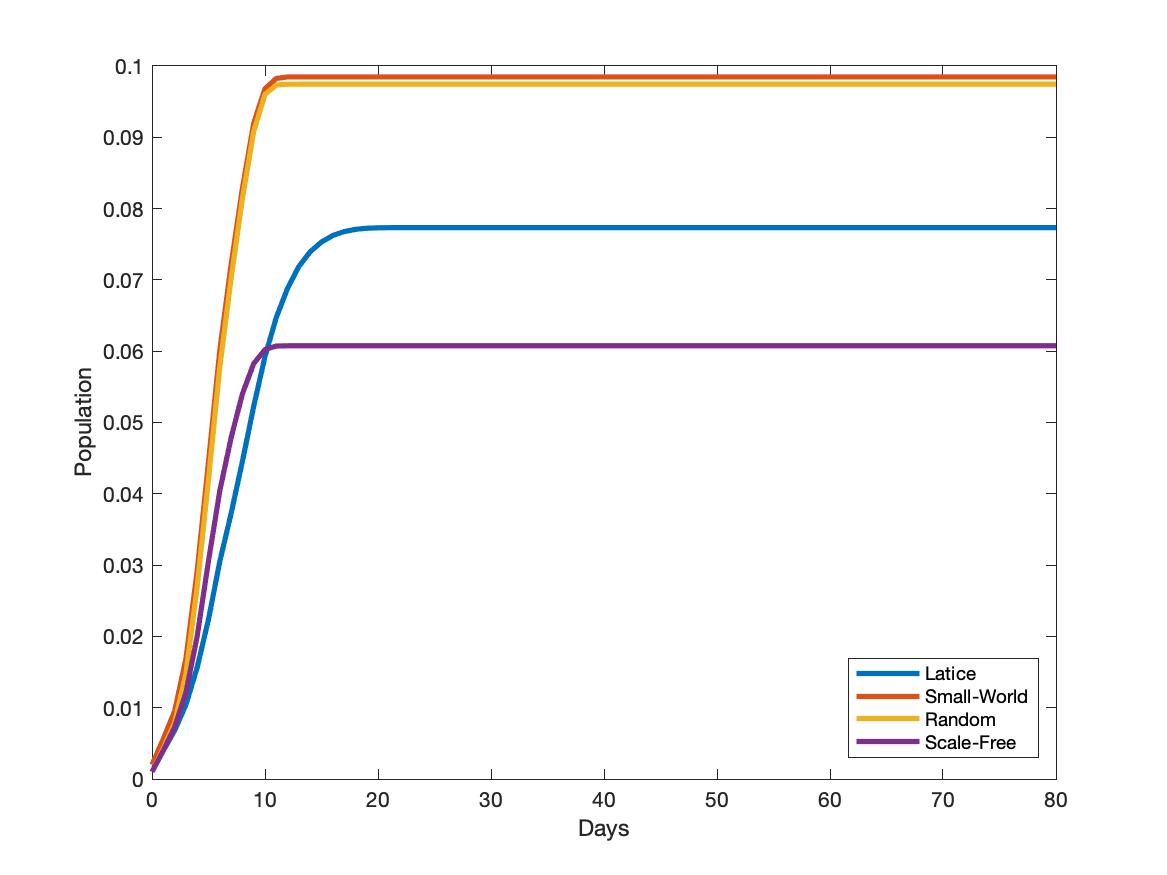}
\caption{Dynamics of Lockdown}\label{lockdownsnet}
\end{subfigure}
\begin{subfigure}{0.5\textwidth}
\centering
\includegraphics[width=1\linewidth]{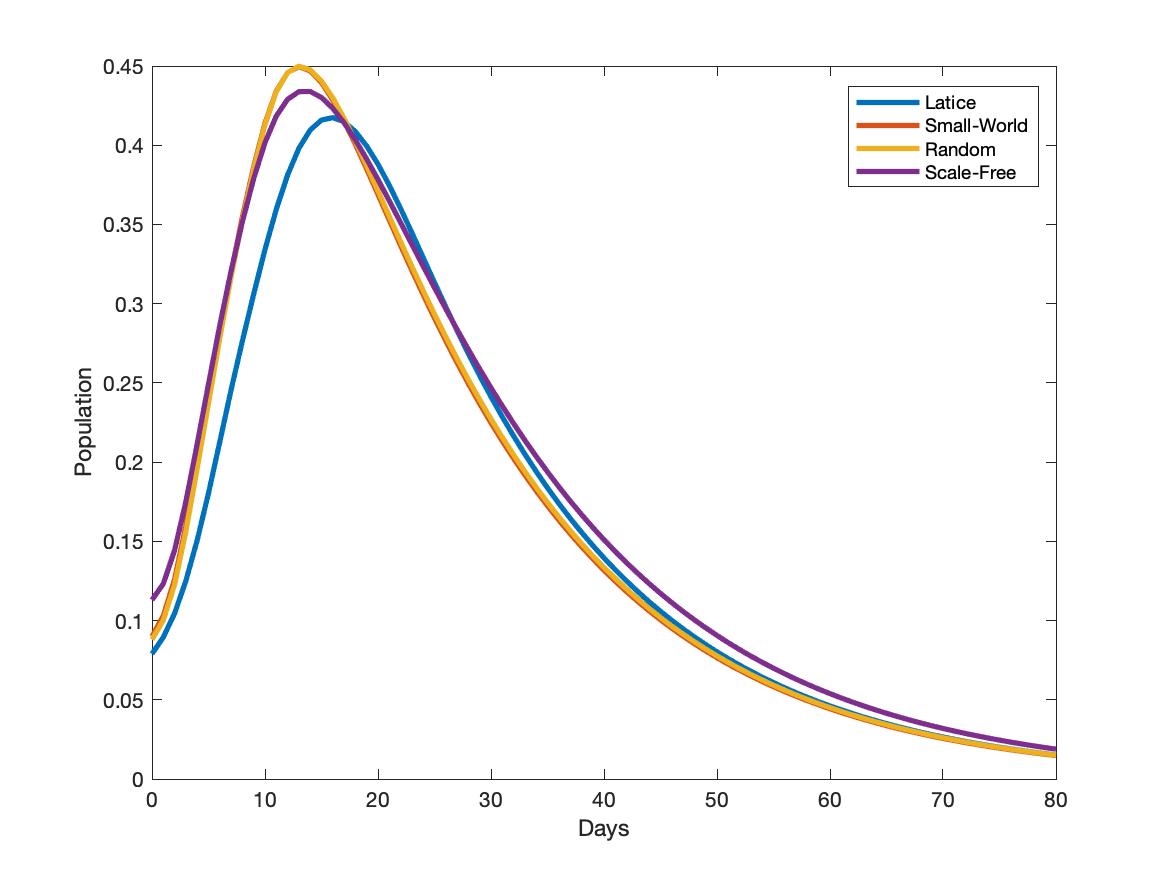}
\caption{Dynamics of Infection}\label{infectionnet}
\end{subfigure}
\begin{subfigure}{0.5\textwidth}
\centering
\includegraphics[width=1\linewidth]{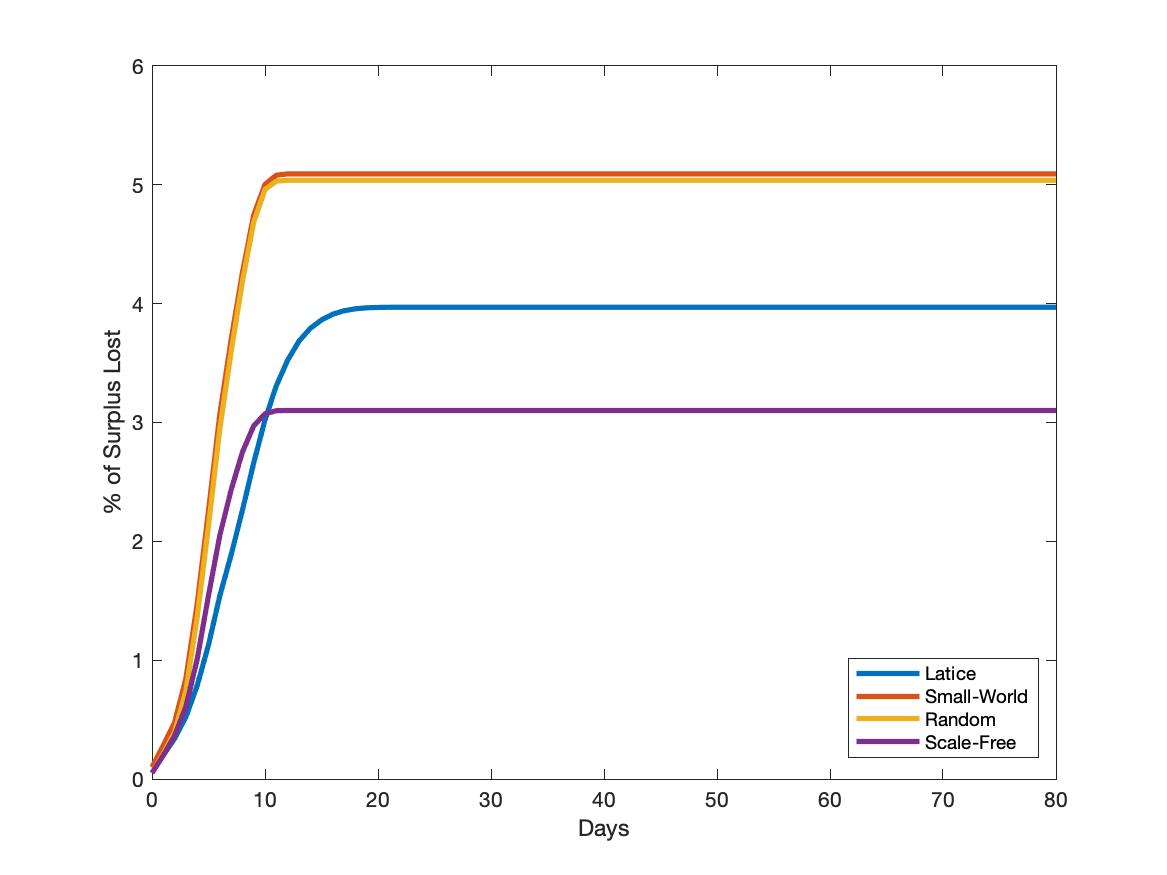}
\caption{Dynamics of Economic Cost}\label{costscnet}
\end{subfigure}
\begin{subfigure}{0.5\textwidth}
\centering
\includegraphics[width=1\linewidth]{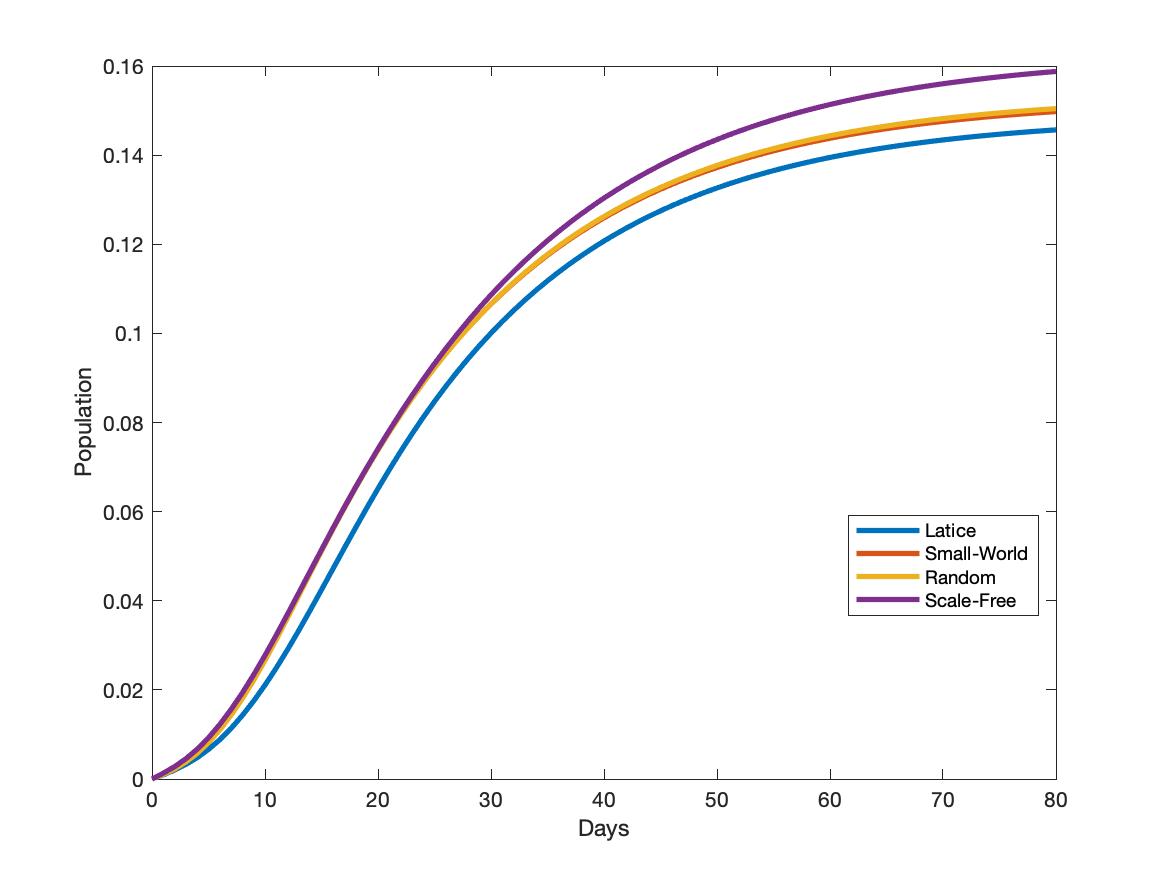}
\caption{Dynamics of Death}\label{deathnet}
\end{subfigure}
\caption{\textbf{Optimal Disease and Economic Costs Dynamics in Networks.} \small{N-SIRD epidemics, lockdown, and economic cost dynamics on the four network types shown in Figure \ref{net_struc}. Each graph shows four curves corresponding to four networks for a single variable of interest. All variability within each curve in each graph is a result of the stochastic nature of transmission and not variation in the network. In the simulation, we assume that the tolerable infection incidence $\lambda =0.01$. The results are displayed in a two-dimensional graphic, with days in the horizontal axis, and the percentage of population affected for the variable (infection, lockdown, or death) illustrated on the vertical axis. In each day, a point in the graphic represents the average value of individual probabilities. For the economic cost, the vertical axis represents the percentage of economic (or surplus) loss relative to the economy without  the pandemic. Based on the simulation results (Figure 4 in the Supplemental Materials) that we obtain by replicating Figure \ref{lock_dyn_sw} with the COVID-19 Delta variant in the small-world network, we conjecture that a replication of Figure \ref{lock_dyn_net} would yield qualitatively consistent results.}}\label{lock_dyn_net}
\end{figure}

We represent the simulation results in the idealized networks in Figure \ref{lock_dyn_net}.

A direct observation is that the epidemic dynamics and the variation in economic costs are different for each society. However, the illustrations in the random network are pretty similar to those in the small-world network. We can explain this similarity by the fact that short path lengths characterize both small-world and random networks. We illustrate the respective optimal lockdown policies in Figure \ref{lockdownsnet} for these four societies. The cumulative proportion of the population sent into lockdown peaks and flattened much earlier in the scale-free network society than in the lattice and small-world networks. At the onset of the pandemic, the lockdown is slightly stricter in the scale-free network compared to the lattice network. However, lockdown is always higher in random and small-world network configurations compared to lattice and scale-free  structures. 

The lockdown dynamics in Figure \ref{lockdownsnet} respond to the disease dynamics that we illustrate in Figure \ref{infectionnet} for infection, and Figure \ref{deathnet} for deaths. We observe that the reduction in initial growth in infection is stronger in lattice networks compared with other networks because high spatial clustering of connections drives a more rapid saturation of local environment \citep{keeling2005networks}. In addition, findings from theoretical models of disease spread through scale-free-network societies show that the infection is generally concentrated among agents with the highest number of connections \citep{pastor2001epidemic, newman2002spread, chang2021mobility}. Therefore, sending these potential super-spreaders into lockdown can significantly reduce the contagion. Our optimal lockdown policy is consistent with these recommendations by suggesting isolating hubs or super-spreaders. Once they are in lockdown in the scale-free network, the speed of infection from one individual to another is reduced (a simple example is a situation in which agents are connected through a star network). The situation is different in the small-world and random network societies, where short path lengths suggest a rapid spatial spread of disease. Then, containing the contagion below a chosen infection incidence level requires drastic lockdown measures. As the epidemic continues, the dynamics of surplus loss that we represent in Figure \ref{costscnet}, due to the pandemic, are different across the four societies, with random and small-world networks suffering the most from the lack of economic activities resulting from significant lockdown. However, the lowest lockdown in scale-free network (Figure \ref{lockdownsnet}) results in more infection and deaths in the long run (Figure \ref{deathnet}).\footnote{Based on the simulation results (Figure 4 in the Supplemental Materials) that we obtain by replicating Figure \ref{lock_dyn_sw} with the COVID-19 Delta variant updated information, we conjecture that a similar exercise with the lattice, random, and scale-free network structures would yield consistent results.}

\begin{figure}[H] 
\begin{subfigure}{0.5\textwidth}
\centering
\includegraphics[width=1\linewidth]{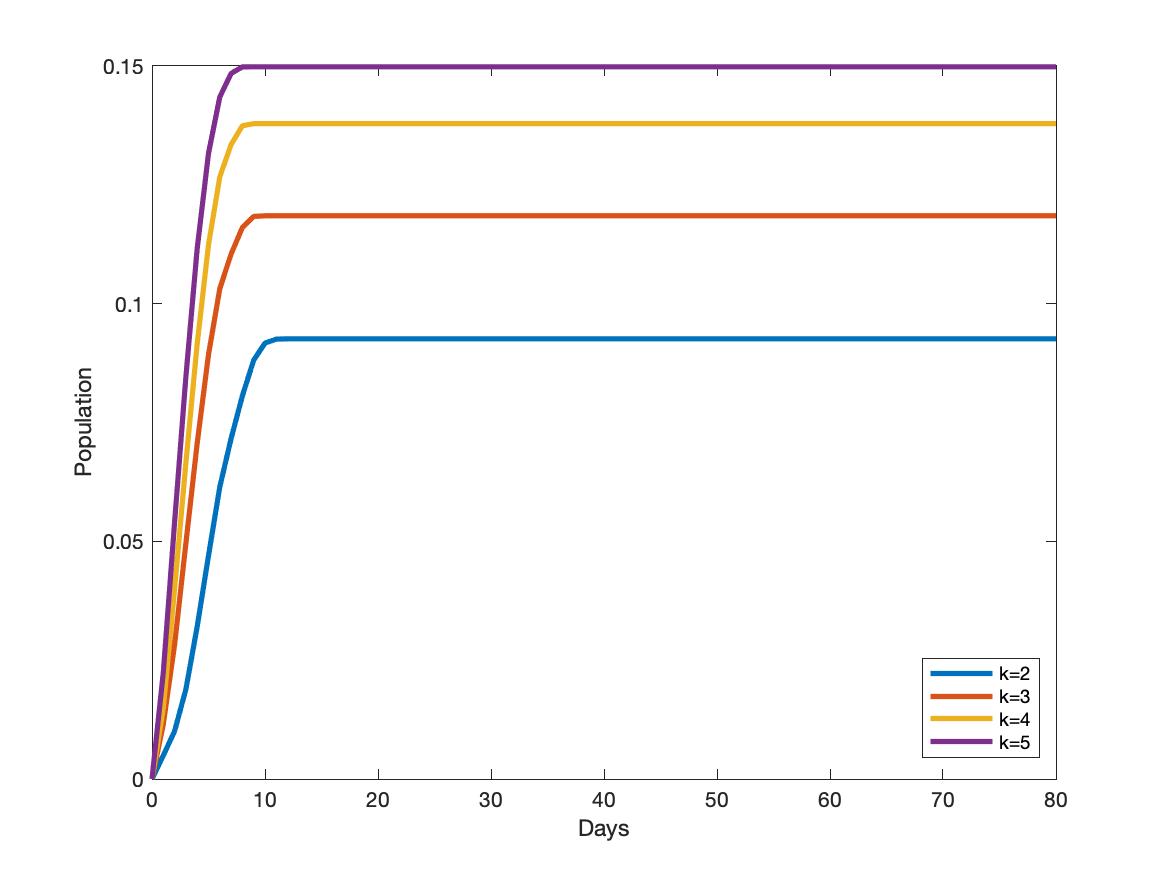}
\caption{Dynamics of Lockdown}\label{lockdownden}
\end{subfigure}
\begin{subfigure}{0.5\textwidth}
\centering
\includegraphics[width=1\linewidth]{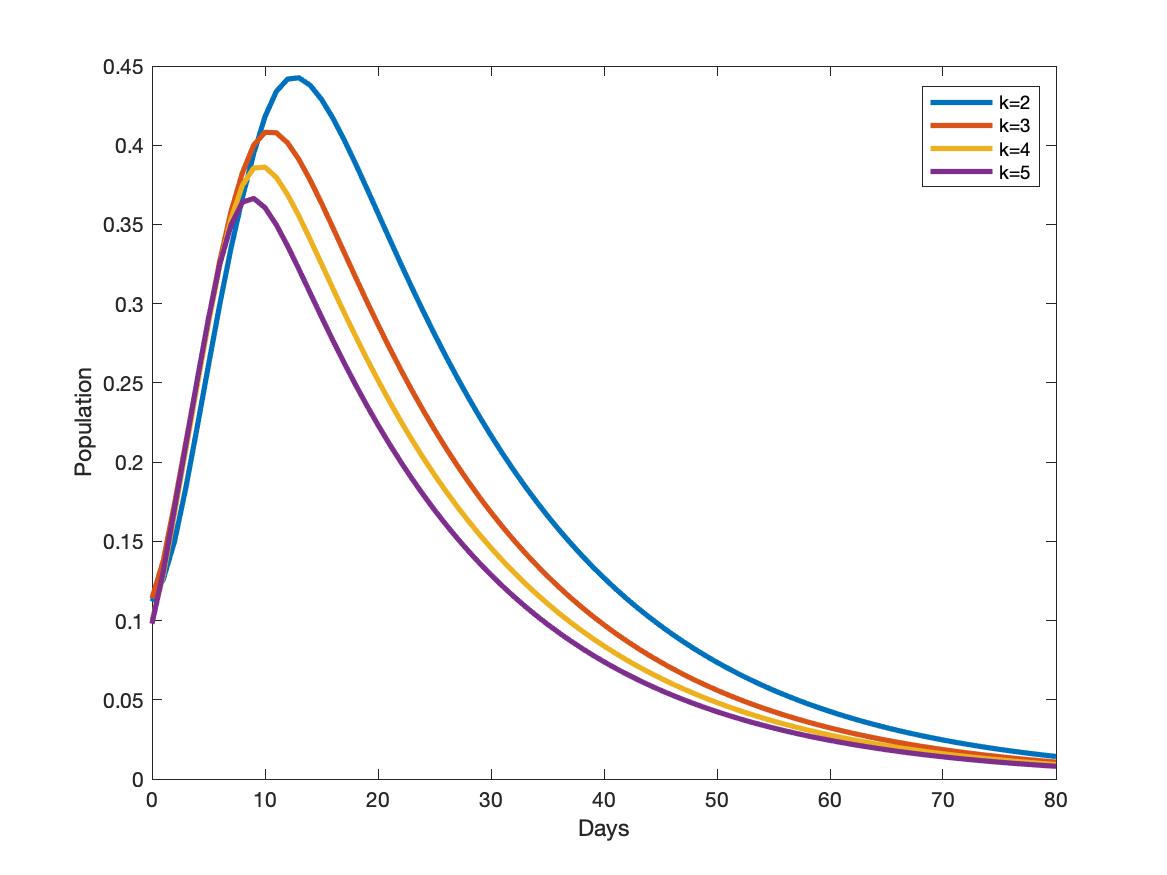}
\caption{Dynamics of Infection}\label{infectionden}
\end{subfigure}
\begin{subfigure}{0.5\textwidth}
\centering
\includegraphics[width=1\linewidth]{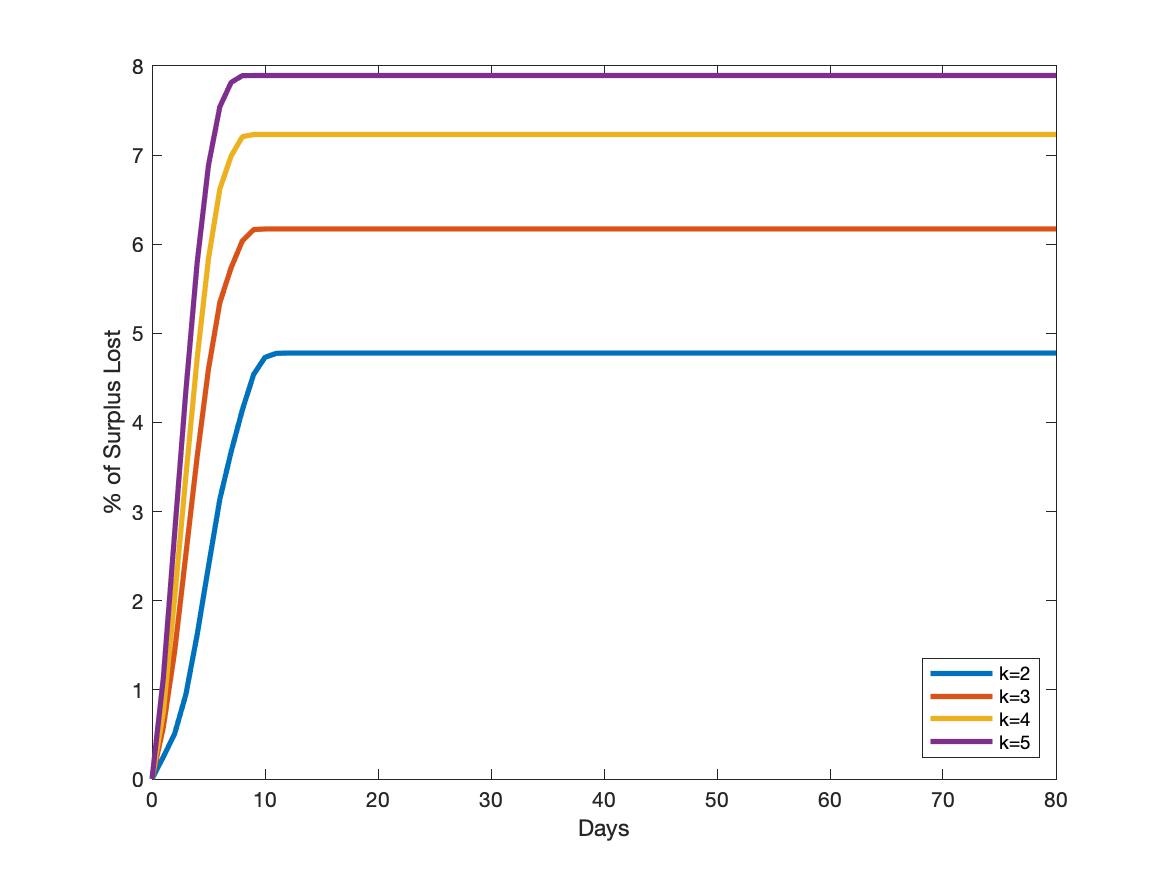}
\caption{Dynamics of Economic Cost}\label{costscden}
\end{subfigure}
\begin{subfigure}{0.5\textwidth}
\centering
\includegraphics[width=1\linewidth]{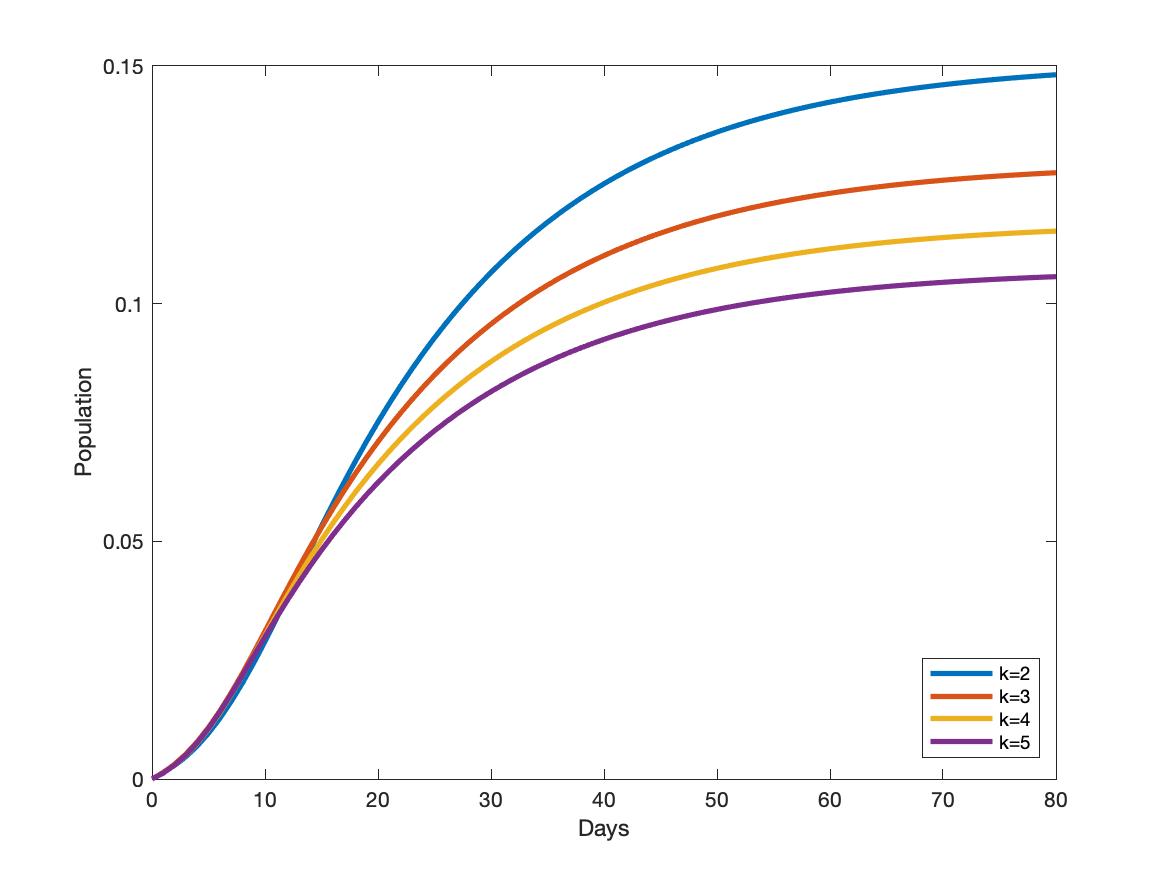}
\caption{Dynamics of Death}\label{deathden}
\end{subfigure}
\caption{\textbf{Optimal Disease and Costs Dynamics in a Small World Networks with Different Densities}. \small{In our simulations, we assume that $\lambda =0.01$. The results are displayed in a two-dimensional graphic, with days in the horizontal axis, and the percentage of population affected for the variable (infection, lockdown, or death) illustrated on the vertical axis. In each period, a point in the graphic represents the average value of individual probabilities. For the economic cost, the vertical axis represents the percentage of economic (or surplus) lost relative to the economy without  the pandemic. The density of network $A_k$ is $d(A_k)= \frac{2k}{n-1}$, where the parameter $k$ represents the average number of connections per agent in network $A_k$, and $n$ the number of nodes. Based on the simulation results (Figure 4 in the Supplemental Materials) that we obtain by replicating Figure \ref{lock_dyn_sw} with the COVID-19 Delta variant in the small-world network, we conjecture that a replication of Figure \ref{lock_dyn_den} would yield qualitatively consistent results.}}\label{lock_dyn_den}
\end{figure}

Following the comparative statics analyses on network topology described in Figure \ref{lock_dyn_net}, one might be interested in knowing how \textit{network density} could affect the optimal lockdown policy, and therefore, the disease dynamics. To address this concern, we consider a society, $A_k$, consisting of $n=1000$ agents connected through a small-world network \citep{watts1998collective} with  $k\times n$ edges, where $k$ represents the average number of connections per agent in the society. The density $d(A_k)$ of the network $A_k$ measures how many ties between agents exist compared to how many ties between agents are possible, given the number of nodes, $n$, and the number of edges, $k\times n$. Since $A_k$ is an undirected network,  $d(A_k)= \frac{2k}{n-1}$, and the network is \textit{dense} (i.e., there is a lot of interconnections between agents) as $k$ increases. Figure \ref{lock_dyn_den} represents the simulation results in society $A_k$, when $k \in \{2, 3, 4, 5\}$. The optimal lockdown dynamics displayed in Figure \ref{lockdownden} illustrate that lockdown probabilities increase with network density. The social planner justifies the latter decision by the fact that the infection rate is slightly high in more dense societies at the onset of the pandemic, as portrayed in Figure \ref{infectionden}. As the pandemic evolves, strict lockdown is effective in containing the infection so that in the long run, less dense societies bear a higher number of deaths relative to more dense societies in Figure \ref{costscden}. Similar to Figure  \ref{lock_dyn_net}, fewer economic transactions as a result of rigid lockdown in more dense networks induce significant surplus loss as displayed in Figure \ref{costscden}.

Our simple experiment in Section \ref{illustration1} sufficiently highlights the fact that network configuration should be a key factor in designing optimal lockdown policies during a pandemic like COVID-19, and this has implications for health dynamics and economic costs. Indeed, our illustrations are  consistent with other studies showing that network configuration plays an essential role in the level of infection spread and information diffusion (see, for example, \citet{keeling2005networks}, \citet{pongou2013fidelity}, \citet{pongou2016volume}, and recently,  \cite{Kuchler_covidfacebook2020},  \cite{Harris2020}, \citet{chang2021mobility}, and \citet{bubar2021model}, among others). The numerical analysis also suggests that the wide range of variation in COVID-19 outcomes observed across countries and across communities within countries could be explained by differences in their network configuration. Several studies analyze the differences in COVID-19 outcomes between countries worldwide and communities within countries or regions. For comparisons among countries, we can cite among others, \citet{balmford2020cross}, \citet{banik2020covid}, \citet{sorci2020explaining}, \citet{abu2020factors}, \citet{schellekens2020unreal}, \citet{karanikolos2020comparable}, and \citet{rice2021variation}. For cross-communities comparisons in COVID-19 outcomes in the United States, see for example, \citet{wadhera2020variation},  \citet{adhikari2020assessment}, \citet{chang2021mobility}, and \citet{hong2021exposure}.

\subsection{Network Centrality and Optimally Targeted Lockdown}\label{centralitylockdown}

\ \ \ \ Our third comparative statics analysis highlights how lockdown policies can be optimally targeted at individuals based on their characteristics. The individual characteristic we consider is centrality in the contact network. In general, in a networked economy, certain agents occupy more central positions than others in the prevailing contact network \citep{albert1999diameter, barabasi1999emergence, jeong2000large, liljeros2001web, chang2021mobility}. This can be due to a variety of reasons, including the distinct social and economic roles played by each individual. It is argued that individuals who occupy more central positions in networks are more likely to be infected and to spread an infection \citep{anderson1992infectious, pastor2001epidemic, newman2002spread, hethcote2014gonorrhea, pongou2018valuing,  rodrigues2019network}. This suggests that an optimal lockdown policy should be targeted at more central agents in a network. However, various measures of network centrality exist, and it is not clear which of these measures are more predictive in the context of a pandemic like COVID-19. 

To address this issue, we consider four popular network metrics: degree centrality, eigenvector centrality, betweenness centrality, and closeness centrality. For clarity, we briefly define these four network centrality measures. \footnote{We should stress that network centrality is not an element in the planning problem's optimization. We provide the analysis to illustrate additional features of our N-SIRD model.} We recall that the network $A$ is a symmetric $n\times n$ weighted adjacent matrix $(A_{ij})$. An agent's \textit{degree centrality} $\chi_i$ equals the total number of other agents directly connected to agent $i$ (i.e., the number of agent $i$'s neighbors): 
\begin{equation*}
\chi_i= \sum\limits_{j=1}^{n} A_{ij}.    
\end{equation*}
\textit{Eigenvector centrality} $\nu_i$ measures the extent to which agent $i$ is connected to other highly connected agents in the network $A$: 
\begin{equation*}
\nu_i = \frac{1}{e} \sum\limits_{j=1}^{n} A_{ij} \nu_j.    
\end{equation*}
The eigenvector centrality is computed using the principal eigenvector $e$ of the adjacent matrix $A$, that we can write in matrix notation as $A \nu = e \nu$, where $\nu$ is a column vector with $n$ entries. The eigenvector centrality reflects the notion that connections to high connected agents  are more important. Agent $i$'s betweenness centrality, $b_i$, measures the fraction of shortest paths passing through agent $i$:
\begin{equation*}
b_i=\sum\limits_{j, k} \frac{\sigma_{jk}^{i}}{\sigma_{jk}},    
\end{equation*}
where $\sigma_{jk}$ is the total number of shortest paths from agents $j$ to $k$, and $\sigma_{jk}^{i}$ is the number of those paths that pass through agent $i$. Agent $i$'s  \textit{closeness centrality}, $c_i$, measures how close is the agent to all other agents in the network $A$:
\begin{equation*}
c_i= \frac{1}{\sum\limits_{j}d(i, j)},    
\end{equation*}
where $d(i, j)$ is the distance (or shortest path) between agents $i$ and $j$. It follows from these definitions that the degree centrality is less based on network configuration than the other centrality measures. To answer the question of how each of the aforementioned network metrics predicts the probability of lockdown, we consider a society in which agents are connected through a small-world network with  $2\times n$ edges. They occupy very distinct positions in this network, resulting in some agents being more central than others. For robustness, our simulation analysis assumes three different values for the tolerable infection incidence $\lambda$:  0.01, 0.05, and 0.1.
\begin{table}[!h]
  \centering
  \begin{tabular}{c c c c c c c c c }
    \toprule
    \multicolumn{1}{l}{$\lambda$} & \multicolumn{2}{c}{Degree ($\chi_i$)} & \multicolumn{2}{c}{Closeness ($c_i$)} & \multicolumn{2}{c}{Betweness ($b_i$)} & \multicolumn{2}{c}{Eigenvector ($\nu_i$)} \\
    \midrule
          & \multicolumn{1}{c}{corr} & \multicolumn{1}{c}{$p$-value} & \multicolumn{1}{c}{corr} & \multicolumn{1}{c}{$p$-value} & \multicolumn{1}{c}{corr} & \multicolumn{1}{c}{$p$-value} & \multicolumn{1}{c}{corr} & \multicolumn{1}{c}{$p$-value} \\
    \midrule
    0.1   & \textbf{0.36} & 8e-33 & \textbf{0.34} & 9e-29 & \textbf{0.33} & 3e-27 & \textbf{0.29} & 1e-20\\
    \midrule
    0.05  & \textbf{0.25} & 5e-16 & \textbf{0.21} & 1e-11 & \textbf{0.21} & 6e-12 & \textbf{0.17} & 1e-07\\
    \midrule
    0.01  & \textbf{0.26} & 1e-16 & \textbf{0.18} & 4e-09 & \textbf{0.18} & 3e-09 & \textbf{0.13} & 4e-05\\
    \bottomrule
    \end{tabular}
\caption{\textbf{Correlation between measures of centrality and average optimal lockdown probability in a small-world network}. \small{In the Supplemental Materials (see Table 4), we replicate Table \ref{tab_central} for scale-free, random, and lattice networks. We also replicate Table \ref{tab_central} using recent epidemiological data of the COVID-19 Delta variant (see Table 5 in the Supplement Materials). We find that the simulation results in Table 3 are qualitatively consistent with the findings in Table  \ref{tab_central}. We conjecture that a similar replication of Table \ref{tab_central} with updated COVID-19 information in lattice, random, and scale-free network would yield consistent qualitative results as in Table 4. The $p$-value for each centrality measure is for the test of the hypothesis $H_0$ $\rho=0$ vs $H_1$ $\rho \neq 0$.}}
  \label{tab_central}%
\end{table}

Table \ref{tab_central} reports the correlation between each of our network metrics and the average optimal lockdown probabilities for different values of the tolerable infection incidence, $\lambda$, in a small-world network. Our simulation results in Table \ref{tab_central} suggest that the four centrality measures positively correlate to the likelihood of lockdown under the optimal lockdown policy. This correlation is statistically significant, as implied by the different $p$-value statistics. Moreover, the predictive force of lockdown obtained for each  centrality measure increases with larger values of $\lambda$.

In Table 4 in the Supplemental Materials, we provide for robustness check other correlations between the four network metrics and average optimal lockdown probabilities for lattice, random, and scale-free networks. We observe that all other centrality measures are positively correlated with the average optimal lockdown probabilities apart from the lattice network. Also, in line with the small-world network, the degree centrality appears to have a stronger correlation with the lockdown in the random and scale-free networks. Though the correlation between the network metrics and optimal lockdown probabilities becomes stronger as the tolerable infection incidence increases in small-world and scale-free networks, the direction of the changes is non-monotonic in lattice and random networks. The latter simulation results suggest that we should be cautious about making any conclusions about the sign and the direction of the relationship between the tolerable infection incidence, $\lambda$, the network centrality measures, and the optimal lockdown probabilities. Nevertheless, the simulation results in Table \ref{tab_central} and in the Supplement Materials (Tables 4 and 5) imply that in a society organized as either a small-world network or a scale-free network, with a higher level of tolerance for the virus, more central individuals will suffer fewer death as a result of being more severely locked. In Section \ref{sec:empiricalApp}, we use data from the network of U.S. nursing and long-term care homes \citep{chen2021nursing} to test some of these simulation results.

Intuitively, though a complete lockdown might be a solution in a pure epidemiological model, it cannot be optimal in our N-SIRD model because the goal is to disconnect the contact network while maintaining economic activities. It follows that under our optimal lockdown policy, not all agents would be in the lockdown. This analysis highlights the limitations of quasi-universal lockdown policies such as those implemented in several countries worldwide in the early period of COVID-19. Our policy recommendation is consistent with studies and reports suggesting shutting down only particular sectors in society during a pandemic like COVID-19 \citep{Acemoglu2020, ipwsj, bosi2021optimal, chang2021mobility, farsalinos2021improved}. These are social and economic hubs, sectors that attract large numbers of individuals, such as large shopping centers, airports and other public transportation infrastructures, schools, certain government buildings, entertainment fields, parks, and beaches, among others. 



\section{\large{Empirical Application: Estimation of State Level Tolerable COVID-19 Infection Incidence using U.S Nursing and Long-Term Care Homes Network}}\label{sec:empiricalApp}
In this section, we calibrate our N-SIRD model, estimate U.S. state level tolerable COVID-19 infection incidence, and test some of the model predictions using unique data from the networks of the U.S. nursing and long-term care facilities. The senior population (adults 65 and older) accounts for a significant share of COVID-19 deaths in the U.S. As of September 24, 2021, seniors account for 16\% of the U.S. population but 77.9\% of U.S. COVID-19 deaths in the U.S. \citep{statista2}. Nursing and long-term care facilities have been at the center of many COVID-19 outbreak, and this situation led the U.S. federal government to instate a ban to nursing home visits on March 13, 2020. This restriction has enabled researchers from the ``Protect Nursing Home Project" to construct a U.S. nursing homes network, using device-level geolocation data for 501,503 smartphones in at least one of the 15,307 nursing homes.  We first use these unique network data in conjunction with nursing home and U.S. state level information to calibrate the N-SIRD model proposed in Sections \ref{N-sirmodellockdown} and \ref{infectiondynamicandlockdown}. The main constraint introduced in our model, the tolerable infection incidence level, $\lambda$, is estimated for each U.S. state using a simulated minimum distance estimator \citep{forneron2018abc,smith1993estimating, gertler1992quality}. The parameter $\lambda$ reveals the extent to which policymakers in different U.S. states are willing to curb the spread of SARSCOV-2, the virus that causes COVID-19. In order words, $\lambda$ captures the planner's tradeoff between health and wealth. A high value for $\lambda$ is equivalent to a ``laissez-faire" situation, indicating a planner's inclination to maximize economic gains even if this theoretically results in more infections and deaths. 

In a cross-sectional model, we estimate the values of $\lambda$ for each U.S state. Using these values, we test some predictions of the N-SIRD model. Precisely, we explore whether the simulation results are consistent with reality. For instance, we examine whether ``laissez-faire" leads to more deaths. We also investigate the effects of network centrality and the tolerable infection incidence on death rates in nursing facilities. Furthermore, given that COVID-19 responses have been highly politicized in the U.S., we investigate how political ideology (measured by the party of the governor) determines $\lambda$. In a recent study, \cite{neelon2021associations} suggest that there is an association between a governor's party affiliation and COVID-19 infections and deaths (see, for example, \citet{baccini2021explaining}, \cite{frankel2021virus}, and \cite{chen2021relationship} for additional evidence linking political party of leaders and COVID-19 fatalities). We complement these findings by investigating the association between the estimated tolerable infection incidence and the governor's party affiliation.

\subsection{Data, Calibration and Estimation of the Tolerable Infection Incidence ($\lambda$)}

To calibrate our parameter of interest, we use data from the Bureau of Labor Statistics, the Senior Living project\footnote{We gather the information from the website: \url{https://www.seniorliving.org/nursing-homes/costs/} consulted on 9/9/2021.}, and the cross-sectional nursing and long-term care homes data provided by \cite{chen2021nursing} for the economic parameters. We obtain the calibration of the epidemiological parameters from Statista.\footnote{The data are available on the web page \url{https://www.statista.com}. The data include the rate of infection or cases, death of COVID-19 among nursing home residents in each U.S. state in September 2020. The page also provides an estimation of the reproduction number of COVID-19 by U.S. state.} Using economic and epidemiological data for each U.S. state, we consider each nursing home as a node in our transmission network. Two nursing homes are connected if the same smartphone signal is recorded in both homes' locations. The number of distinct signals recorded gives a weight to the connection or link between two nursing homes. Nursing and long-term care facilities display a wide range of connectedness with other facilities. \citet{chen2021nursing} use different network metrics to predict COVID-19 in nursing homes. In this empirical application of our N-SIRD model, we focus on the eigenvector centrality, $\nu_i$, which measures the extent to which a nursing home $i$ in a U.S. state is connected to other highly connected nursing homes in the network of nursing facilities in the state.\footnote{In the Supplement Materials (see Table 6), we show that our main empirical results are robust when replacing eigenvector centrality by the degree centrality.} To illustrate how the eigenvector centrality measure differs across nursing homes, we present network graphs for a subset of homes in six states as depicted in Figure \ref{net_struc} and summarized in Table 3 in the Supplement Materials.\footnote{We use the algorithm by \citet{fruchterman1991graph}  and the igraph R package \citep{csardi2006igraph} to plot our selected six network structures.} More-connected nursing facilities are generally toward the center of each graph, and facilities with fewer contacts are on the periphery. Table \ref{nurstasts} summarizes the descriptive statistics of U.S. nursing and long-term care facilities. We refer to \citet{chen2021nursing} for additional details on nursing homes characteristics and network centrality measures in these care facilities.

\begin{figure}[H] 
\includegraphics[width=1\linewidth]{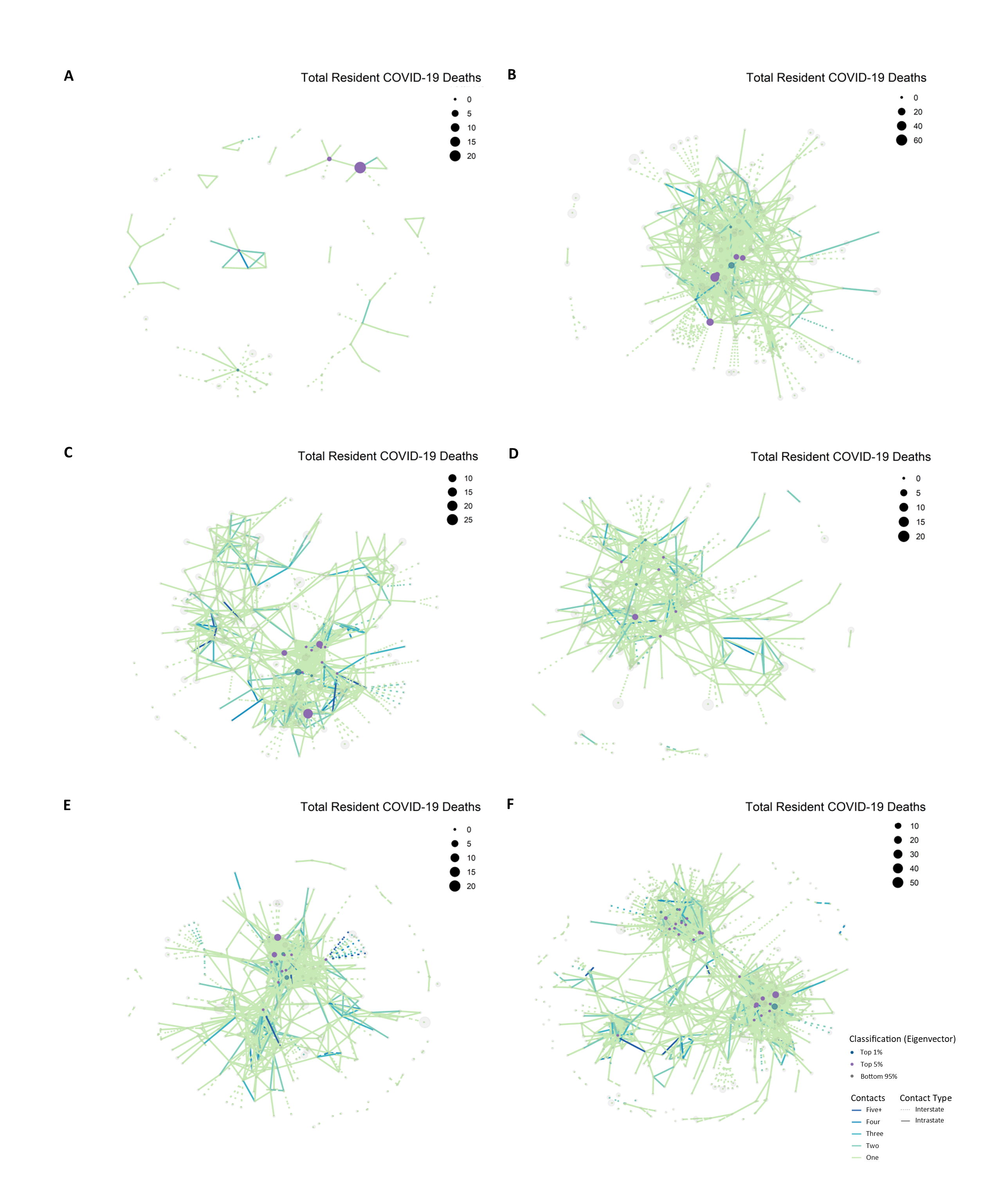}
\caption{\textbf{Network structure of selected nursing home facilities} in South Dakota (A), Connecticut (B), Louisiana (C), Colorado (D), Oklahoma (E), and Missouri (F). Details for each network structure are provided in Table 3 in the Supplement Materials. }\label{net_struc}
\end{figure}

\begin{table}[H]
\centering
\begin{tabular}{rll}
  \hline
  & Variable & State Reporting Facilities \\ 
  \hline
  & Number of Nursing Homes & 15277 \\ 
  & \textbf{COVID-19 information} \\
  & Cases & 84.47 (237) \\ 
  & Death & 1.84 (5.94) \\ 
  & \textbf{Network metrics} \\
  & Home Degree & 6.21 (7.83) \\ 
  & Home Eigenvector Centrality & 0.08 (0.18) \\ 
  & \textbf{Regulatory measures} \\
  & For Profit, \% & 70.3 \\ 
  & Number of Beds & 105.61 (59.04) \\ 
  & Number of Beds Occupied & 76.97 (48.01) \\ 
  & CMS quality rating (1-5) & 3.69 (1.24) \\ 
  & County SSA & 391.39 (273.53) \\  
   \hline
\end{tabular}
\caption{\textbf{Descriptive statistics of nursing homes from \cite{chen2021nursing}'s Replication data}. \small{Binary variables are percent of nursing homes; continuous variables are mean values, with standard deviations in parentheses. In our regression-based analysis, we only use the information on the nursing homes in the 26 U.S. states in which we estimate the tolerable COVID-19 infection incidence. This sample consists of 6478 observations, which is still a sufficiently large sample for our empirical exercise.}}\label{nurstasts}
\end{table}

The parameters of the model that we propose in Sections \ref{N-sirmodellockdown} and \ref{infectiondynamicandlockdown} are obtained for each U.S. state by following the calibration method that we described in Table \ref{tab:addlabel}.

\begin{landscape}
\begin{table}[htbp]
  \centering %
\resizebox{20cm}{7cm}{\begin{tabular}{ l c l}
\cmidrule{1-3}    \textbf{Parameters} & \textbf{Value} & \textbf{Definitions and Sources}   \\
\cmidrule{1-3}    \textbf{Epidemiological} &       &         \\
\cmidrule{1-3}    $\beta$ & $R_0/18$ & The reproduction numbers $R_0$ estimate April-July 2020, from Statista.  \\
\cmidrule{1-3}    $\gamma$ & (1-death/case)/18 & case and death per 1000 for in nursing home by U.S. state in  Sep. 2020 from Statista   \\
\cmidrule{1-3}    $\kappa$ & (death/case)/18 & case and death per 1000 for in nursing home by state in  Sep. 2020 Statista  \\
\cmidrule{1-3}    Death Count & 80\% of COVID-19 death  &  New York Times Data for each U.S. state \citep{conlen2021more} from May 31 to August 16, 2020   \\
\cmidrule{1-3}    $A$    & Network of Nursing Home & Protect Nursing Home Project   \\
\cmidrule{1-3}    \textbf{Economic} &       &        \\
\cmidrule{1-3}    Price ($p$) & Average hourly cost of a Private Room &   Senior Living Project\footnote{\url{https://www.seniorliving.org/nursing-homes/costs/}} \\
\cmidrule{1-3}    Wage ($w$)  & Average hourly wage by U.S. state & BLS     \\
\cmidrule{1-3}    $\alpha$ & Elasticity of output wrt. capital & Replication data from \citet{chen2021nursing} Estimation for each U.S. state  \\
\cmidrule{1-3}    $\delta$ & 0.05/365 & Atkeson et al. (2020)  \\
\cmidrule{1-3}    $\phi$ & 0 &  The authors      \\
\cmidrule{1-3}    $\psi$ & 1 &   The authors      \\
\cmidrule{1-3}    Capital ($k$) & Number of beds in the Nursing Home & Replication data from \citet{chen2021nursing} 
 \end{tabular}}%
\caption{\textbf{Calibration of Parameters for each U.S. state}}\label{tab:addlabel}
\end{table}%
\end{landscape}

After calibration, for each U.S. state, there is a parameter remaining, $\lambda$, the parameter measuring the governor's tolerable infection incidence. We estimate $\lambda$ using a simulated minimum distance estimator.  Indeed, for each potential value of  $\lambda$, the planner's problem is solved and the dynamics of death of the model over 77 days is compared with the raw data of elderly death dynamics provided by the New York Times death count from May 31 to August 16, 2020.  The value of $\lambda$ that will minimize the distance between the two dynamics will be the  estimate of the tolerable infection incidence level of the U.S. state's social planner.

The procedure is carried-out for 49 U.S. states present in our data set.  Out of the 49, 26 U.S. states deliver estimates of the tolerable infection incidence level by the policymakers that are significantly different from zero.\footnote{For the remaining U.S. states, the parameter $\lambda$ is not identified as the procedure always return the initial value, suggesting a flat objective function. Indeed, the simulated minimum distance estimator in our sample does not guarantee the identification of $\lambda$ as the raw data (New York Times death count) is not collected exactly on the nursing homes. Moreover, there are conditions on the network matrix necessary to solve the planner's problem that may not be satisfied in some U.S. states.} The estimated tolerable infection incidence level range from 0.0006 for the state of Missouri (MO) to 0.45 for Alabama (AL). The average value of $\lambda$ is 0.12 and the standard deviation is 0.13 indicating a substantial level of dispersion.

\begin{figure}[!h] 
\centering
\includegraphics[width=.8\linewidth]{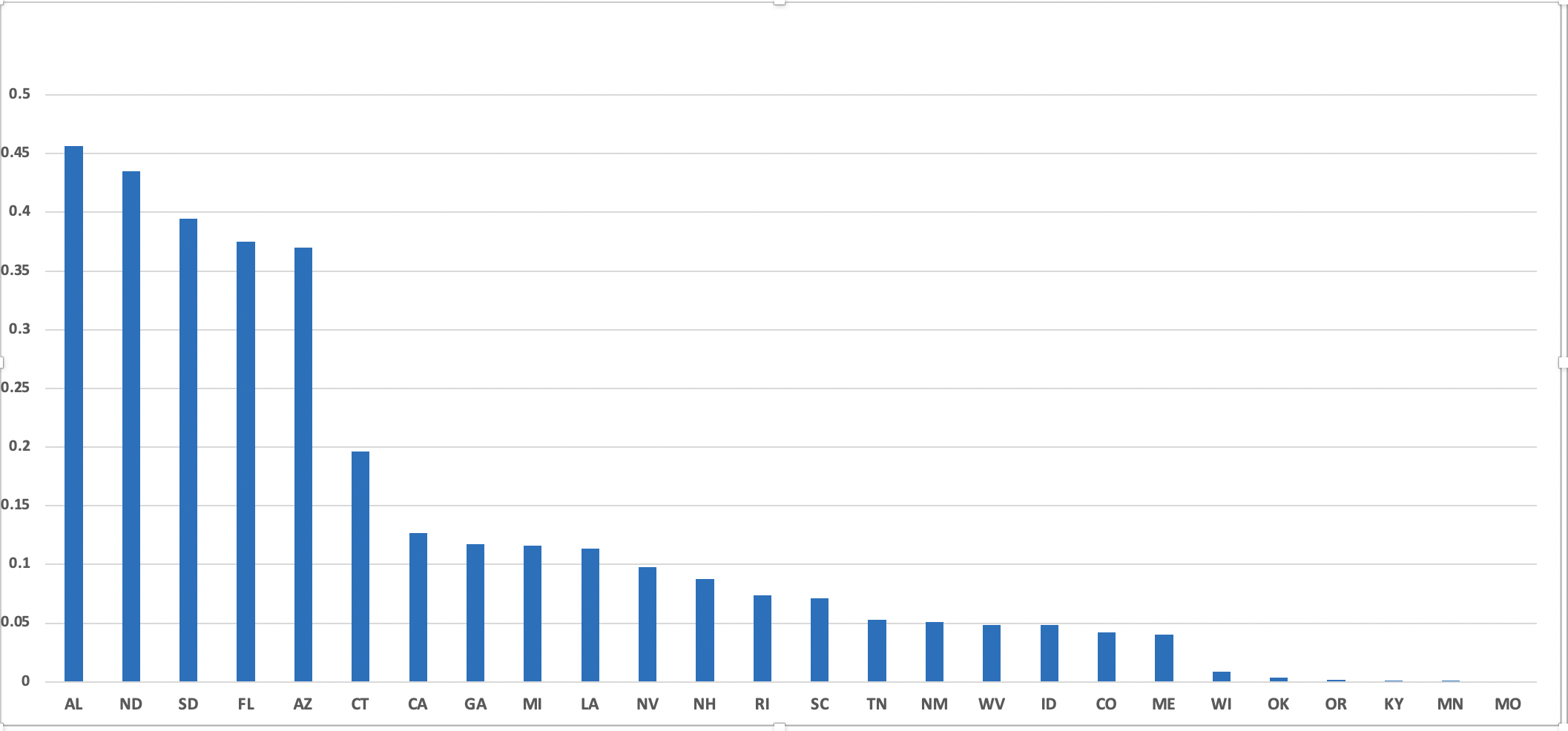}
\caption{\textbf{U.S. state's Tolerable Infection Incidence Level $\lambda$}. \small{The parameter $\lambda$ estimates the tolerable COVID-19 infection incidence of the U.S. state governor from May 31 to August 16, 2020. Using the data and the N-SIRD model with lockdown, we estimate $\lambda$ for 26 U.S. states. The estimates range from 0.0006 for Missouri (MO) to 0.45 for Alabama (AL). The average value of estimates is 0.12 and the standard deviation is 0.13.}} \label{Inc_to}
\end{figure}

\subsection{Testing Some N-SIRD Model's Predictions}\label{sec:testmodel}

Our data sample has the total number of COVID-19 deaths in each nursing home throughout the study (May 31 to August 16, 2020).  We estimate the following simple linear model.
\begin{eqnarray}\label{equ:regm1}
covid\_death_{ijs}  &=& a_0 \lambda_s + a_1  Eig\_Cent_{ijs}+ a_2 County\_ssa_{js} +a_3 D\_Profit_{ijs} \\ \nonumber
                    &+&    b_1 \lambda_s\times Eig\_Cent_{ijs} + b_2\lambda_s\times County\_ssa_{js}+b_3 \lambda_s\times D\_Profit_{ijs} \\ 
                   &+&  c'X_{ijs}+ \theta_j + \varepsilon_{ijs}, \nonumber
\end{eqnarray}
where $covid\_death_{ijs}$ is a variable counting the total number of COVID-19 deaths in the nursing home $i$, in County $j$ and U.S. state $s$,  $\lambda_s $ is the tolerable infection incidence in U.S. state $s$, $Eig\_Cent_{ijs}$ is the eigenvector centrality index for the nursing home, $County\_ssa_{js}$  is the county $j$'s average socio-economic status, $D\_Profit_{ijs}$ is an indicator for whether the nursing home is for profit or not (1 if for-profit, and 0 otherwise), $X_{ijs}$ represents other exogenous characteristics of the nursing home including the constant, and $\theta_j$ is the county fixed effect.\footnote{The choice of the total number of COVID-19 deaths rather than cases, as the outcome variable, is motivated by two reasons. First, the number of COVID-19 cases contains both asymptomatic patients and those who will later recover, so it cannot be an appropriate measure of the human cost of the pandemic. Second, as represented in Figure \ref{incidenceandinfection}, depending on the point in time during the pandemic, there may be no difference in the number of infected individuals as a function of the tolerable infection incidence. On the contrary, the total number of deaths displays an unambiguous dynamics which makes  the theoretical predictions  of the N-SIRD model easier to test. In addition, the total number of deaths is, without doubt, a proxy of the estimate of the human cost of the pandemic.}  The parameters of interest are $a_0$ to $a_3$ and $b_1$ to $b_3$. The estimated values of these parameters can be found in Table \ref{tab:covdeath}.

\begin{table}[htbp]\centering
\def\sym#1{\ifmmode^{#1}\else\(^{#1}\)\fi}
\begin{tabular}{l*{5}{c}}
\hline\hline
                    &\multicolumn{1}{c}{(1)}&\multicolumn{1}{c}{(2)}&\multicolumn{1}{c}{(3)}&\multicolumn{1}{c}{(4)}&\multicolumn{1}{c}{(5)}\\
\hline
$\lambda$             &       0.713\sym{**}  &       1.063\sym{***} &       2.127\sym{***} &      -0.105         &       1.573\sym{**}  \\
                    &      (2.04)         &      (3.19)         &      (3.25)         &     (-0.21)         &      (2.12)         \\
[1em]
Eig\_Cent&       1.006\sym{***} &       1.482\sym{***}&       1.026\sym{***}&       1.011\sym{***} &       1.533\sym{***}\\
                    &      (3.27)         &      (3.98)         &      (3.35)         &      (3.29)         &      (4.11)         \\
[1em]
County\_ssa          &   -0.000780         &   -0.000824         &   -0.000521         &   -0.000793         &   -0.000584         \\
                    &     (-1.09)         &     (-1.15)         &     (-0.72)         &     (-1.11)         &     (-0.81)         \\
[1em]
D\_Profit       &       0.266\sym{**}  &       0.268\sym{**}  &       0.269\sym{**}  &       0.101         &      0.0836         \\
                    &      (2.28)         &      (2.29)         &      (2.30)         &      (0.73)         &      (0.60)         \\
 [1em]
$\lambda \times$ Eig\_Cent           &                     &      -3.944\sym{**}  &                     &                     &      -4.157\sym{**}  \\
                    &                     &     (-1.97)         &                     &                     &     (-2.08)         \\
[1em]
$\lambda \times$ County\_ssa            &                     &                     &    -0.00446\sym{**}  &                     &    -0.00445\sym{**}  \\
                    &                     &                     &     (-2.47)         &                     &     (-2.48)         \\
[1em]
$\lambda \times$ D\_Profit             &                     &                     &                     &       1.231\sym{**}  &       1.387\sym{**}  \\
                    &                     &                     &                     &      (1.97)         &      (2.20)         \\
                    [1em]
Overall\_Rating      &      -0.207\sym{***}&      -0.207\sym{***}&      -0.207\sym{***}&      -0.210\sym{***}&      -0.210\sym{***}\\
                    &     (-5.07)         &     (-5.07)         &     (-5.06)         &     (-5.13)         &     (-5.12)         \\
[1em]
County FE          &       Yes &       Yes  &       Yes &       Yes &      Yes \\
\hline
Observations        &        6478         &        6478         &        6478         &        6478         &        6478         \\
\(R^{2}\)           &       0.072         &       0.073         &       0.073         &       0.073         &       0.074         \\
\hline\hline
\end{tabular}
\caption{\textbf{Estimation of the ``laissez-faire" Effect: the Dependant variable is the Total Number of COVID-19 Deaths in the Nursing Home.} \small{Table \ref{tab:covdeath} summarizes the estimation of the effects of U.S. state's tolerable COVID-19 infection incidence level ($\lambda$)  on the total number of COVID-19 death in U.S. nursing homes. Standard errors  are robust to heteroscedasticity of unknown form.} \footnotesize \textit{t} statistics in parentheses. \footnotesize \sym{*} \(p<0.1\), \sym{**} \(p<0.05\), \sym{***} \(p<0.01\). In the Supplement Materials (see Table 6), we show that our main empirical results in Table \ref{tab:covdeath} are robust when replacing eigenvector centrality by the degree centrality.}\label{tab:covdeath}
\end{table}

Estimating the tolerable infection incidence level in each U.S. state allows us to verify some of the model predictions. Figure \ref{incidenceanddeaths} in Section \ref{tradeoff} illustrates the relationship between the tolerable infection incidence level of the epidemic and the death dynamics. This static comparative implies that a more significant value for $\lambda$ is associated with more COVID-19 deaths.  The OLS estimation results in Table \ref{tab:covdeath} suggest that there is a negative association between the tolerable infection incidence level and the total number of COVID-19 deaths in a nursing home. A five standard-deviation increase in the tolerable infection incidence is expected to lead approximately to one additional death in a nursing home everything else being equal. This means that more ``laissez-faire" will result in more COVID-19 deaths.  This result is robust to controlling for county fixed-effects, the level of income of the resident proxied by the county average socio-economic status, the quality of care provided, and whether the nursing home is or not for profit.

The simulation results uncovered in Section \ref{sec:simulations} suggest that the level of network centrality plays a pivotal role in the optimal lockdown and diffusion of the epidemic. The optimal lockdown policy targets more central individuals with a higher probability. Table \ref{tab_central} in Section \ref{centralitylockdown} suggests that higher values of tolerable infection incidence levels are associated with a higher likelihood of lockdown of central agents in the network structure.  Therefore, our simulation would predict that more ``laissez-faire" (i.e., increase in $\lambda$) will attenuate the impact of network centrality on the number of COVID-19 deaths because more central individuals will be lockdown. In other words, under a laissez-faire policy, the difference in the number of deaths between central and peripheral nursing homes is reduced. The regression results in Table \ref{tab:covdeath} test this prediction.  Column (1) shows that being more central is associated with more COVID-19 deaths in the nursing homes. Column (2) shows the interaction between eigenvector centrality and the tolerable infection incidence. The interaction term has a negative and statistically significant effect on total COVID-19 deaths. An increase in the level of the tolerable infection incidence reduces the relative death toll of more central nursing homes.  Columns (2) and (5) of Table \ref{tab:covdeath} show the robustness of this result to the introduction of several controls.  We also verify another prediction of our model's simulation in the sample under investigation. Our results complement \cite{chen2021nursing} by showing that while the level of eigenvector centrality matters in the propagation of the epidemic and death count, there exists heterogeneity in the extent of its relevance.  More precisely, we show that the social planner's tolerable infection incidence affects the relationship between the level of centrality and the number of COVID-19 deaths. This relationship is less pronounced under a laissez-faire regime.

We also assess how ``laissez-faire" affects the relationship between the COVID-19 death toll and economic conditions and type of nursing home (for a profit or not).  Column (5) in Table \ref{tab:covdeath} shows that laisser-faire more negatively affects nursing homes in economically deprived counties.  Next, we analyze the impact of the nursing home type on the total number of death. Our estimations suggest that for-profit nursing homes have seen 27\% more deaths compared to not-for-profit ones (see Columns (1) to (3) in Table  \ref{tab:covdeath}). Moreover, our results show that the type of the nursing home and the tolerable infection incidence are the main drivers of the difference in COVID deaths in nursing homes. Indeed, when we introduce the interaction term between $\lambda$ and for-profit (D\_Profit) in Column (4), both $\lambda$ and the for-profit indicator (D\_Profit) become smaller in absolute value and non-statistically significant; only the interaction term has a positive and statistically significant coefficient. We also note that better rated nursing homes have significantly less deaths. 

\begin{table}[htbp]\centering
\def\sym#1{\ifmmode^{#1}\else\(^{#1}\)\fi}
\begin{tabular}{l*{5}{c}}
\hline\hline
                    &\multicolumn{1}{c}{(1)}&\multicolumn{1}{c}{(2)}&\multicolumn{1}{c}{(3)}&\multicolumn{1}{c}{(4)}&\multicolumn{1}{c}{(5)}\\
\hline
$\lambda$              &       3.812\sym{**} &       4.290\sym{**} &       5.401\sym{**} &                     &                     \\
                    &      (2.57)         &      (2.15)         &      (2.35)         &                     &                     \\
[1em]
Democrat Governor                 &                     &      0.0554         &       0.845         &                     &      -2.040\sym{**} \\
                    &                     &      (0.08)         &      (1.02)         &                     &     (-2.22)         \\
[1em]
Female Governor              &                     &      -0.513         &      -0.765         &                     &      -0.593         \\
                    &                     &     (-0.91)         &     (-1.31)         &                     &     (-0.96)         \\
[1em]
South               &                     &      -1.083\sym{*}  &      -1.124\sym{*}  &                     &      -1.261\sym{*}  \\
                    &                     &     (-1.78)         &     (-1.92)         &                     &     (-2.06)         \\
[1em]
Democrat $\times$ $\lambda$ &                     &                     &      -10.15\sym{*}  &                     &                     \\
                    &                     &                     &     (-1.95)         &                     &                     \\
[1em]
$log(\lambda)$            &                     &                     &                     &       0.169         &       0.420\sym{**} \\
                    &                     &                     &                     &      (1.47)         &      (2.25)         \\
[1em]
Democrat $\times$ $log(\lambda)$        &                     &                     &                     &                     &      -0.540\sym{**} \\
                    &                     &                     &                     &                     &     (-2.38)         \\
[1em]
Constant            &      -4.100\sym{***}&      -3.674\sym{***}&      -3.804\sym{***}&      -3.088\sym{***}&      -1.700\sym{*}  \\
                    &    (-11.87)         &     (-5.16)         &     (-5.32)         &     (-6.57)         &     (-2.05)         \\
\hline
Observations        &          26         &          26         &          26         &          26         &          26         \\
\(R^{2}\)           &       0.165         &       0.307         &       0.382         &       0.057         &       0.320         \\
\hline\hline
\end{tabular}
\caption{\textbf{Estimation of the ``laissez-faire" Effect on U.S. state's GDP Growth in 2020}. \small Standard errors  are robust to heteroscedasticity of unknown form.  \sym{*} \(p<0.1\), \sym{**} \(p<0.05\), \sym{***} \(p<0.01\), \textit{t} statistics are in parentheses.  In Table 7 in the Supplemental Materials, we present the descriptive statistics of GDP and U.S. states' governorship political affiliation and gender in 2020. }\label{tab:growth}
\end{table}

The simulations in Section \ref{sec:simulations} also show  the relationship between the tolerable incidence and the economic performance. Figure \ref{incidenceandcosts} suggests that more ``laissez-faire" is associated with lower total economic cost.  The estimation results in Table \ref{tab:growth} put that prediction to a test by estimating the effect of the U.S. state's tolerable COVID-19 infection incidence on the level of GDP growth in 2020. In accordance with the theoretical simulations, our estimation results suggest a positive  relationship between $\lambda$ and the GDP growth.  The effect of the tolerable infection incidence on economic growth is larger for republican governors.  These results are robust to the controls for regional differences and for the gender of the governor.

In summary, the findings of Table \ref{tab:growth} validate some essential predictions of the N-SIRD model using data from  the U.S. nursing homes networks. Indeed,  we provide evidence suggesting that a higher tolerable infection incidence is associated with more COVID-19 death. Moreover, centrality plays a pivotal role in optimal lockdown, and more ``laissez-faire" exacerbates its effects. We also show that the tolerable infection incidence seems to mediate the impact of economic variables on the human cost of the pandemic. The existence of a positive correlation between tolerable infection incidence and the economic performance is tested and validated in our sample.

In previous sections, we assume that policymakers decide the optimal lockdown policy according to their desirable tolerable infection incidence levels, $\lambda$. Given its importance in controlling the disease and economic dynamics, investigating the sources of its heterogeneity as described in Figure \ref{Inc_to} is of interest. We perform this exercise in Section \ref{politicaloriginInc}.

\subsection{Political Origins of the Tolerable Infection Incidence Heterogeneity}\label{politicaloriginInc}

Whether it is about public health, economic lockdown, mask mandate, or COVID-19 vaccine, the public debates on COVID-19 have been divided along political lines \citep{adolph2021pandemic, neelon2021associations}.  The extent to which this division has affected the COVID-19 pandemic is at the heart of a new and growing literature. We contribute to this literature by examining whether the party affiliation of the U.S. state's governor predicts the tolerable COVID-19 infection incidence. We regress the tolerable infection incidence level on the party affiliation of the incumbent governor in the period covered by the sample (May 31 to August 16, 2020) and other controls.  The regression elucidates the most critical determinant of the U.S. state's choice of the tolerable COVID-19 infection incidence level. 

\begin{table}[!h]\centering
\def\sym#1{\ifmmode^{#1}\else\(^{#1}\)\fi}
\begin{tabular}{l*{4}{c}}
\hline\hline
                    &\multicolumn{1}{c}{(1)}&\multicolumn{1}{c}{(2)}&\multicolumn{1}{c}{(3)}&\multicolumn{1}{c}{(4)}\\
\hline

Republican Governor                 &    0.0973\sym{***}                 &       0.104\sym{***}&      0.0999\sym{***}&      0.0756\sym{***}\\
                    &          (33.06)            &     (33.28)         &     (31.93)         &     (20.59)         \\
[1em]
Republican $\times$ Covid\_Death           &                     &    -0.00293\sym{***}&    -0.00376\sym{***}&    -0.00351\sym{***}\\
                    &                     &     (-5.72)         &     (-7.12)         &     (-6.84)         \\
[1em]
Covid\_Death &                     &     0.00280\sym{***}&     0.00291\sym{***}&     0.00289\sym{***}\\
                    &                     &     (11.00)         &     (10.48)         &      (9.61)         \\
[1em]
Female Governor               &                     &                     &      0.0536\sym{***}&      0.0693\sym{***}\\
                    &                     &                     &     (11.15)         &     (14.17)         \\
[1em]
South               &                     &                     &                     &      0.0514\sym{***}\\
                    &                     &                     &                     &     (12.61)         \\
[1em]
Constant            &       0.174\sym{***}&      0.0718\sym{***}&      0.0656\sym{***}&      0.0553\sym{***}\\
                    &     (63.55)         &     (63.14)         &     (49.25)         &     (33.09)         \\
\hline

Observations        &        6985         &        6564         &        6564         &        6564         \\
\(R^{2}\)           &       0.128         &       0.138         &       0.158         &       0.183         \\
\hline\hline
\end{tabular}
\caption{\textbf{Estimation of the Governor Party Affiliation Effect: The Dependent variable is the U.S. state's Tolerable COVID-19 infection Incidence}. \small Standard errors  are robust to heteroscedasticity of unknown form.  \sym{*} \(p<0.1\), \sym{**} \(p<0.05\), \sym{***} \(p<0.01\), \textit{t} statistics are in parentheses.} \label{tab:lambdaesti}
\end{table}

Democrat governors have 8\% lower tolerable infection incidence, as shown by the estimation results in Table \ref{tab:lambdaesti}.  Thus, Republican governors are more inclined to implement a ``laissez-faire" policy, which mirrors the traditional pro-market position of the party.\footnote{This statement is more in line with the position of Texas Republican Lieutenant Governor Dan Patrick suggesting that seniors are willing to die to keep the economy afloat ``Those of us who are 70 plus, we'll take care of ourselves. But don't sacrifice the country" \citep[n.d]{knodel2020}. Similarly, in a recent study, \citet{baccini2021explaining} find consistent results about the role of political ideology on the responses to the COVID-19 pandemic. For instance, their results suggest that during the early COVID-19 epidemic, Democratic governors emphasized health and safety and were significantly more likely to implement a statewide order. In contrast, Republican governors were particularly concerned about the economic costs of stay-at-home measures.} The result is robust to controlling for geographical U.S. state characteristics, the severity of the epidemic, and the gender of the state's governor.

Not surprisingly, there is a positive association between the number of deaths in the nursing home and the U.S. state's tolerable COVID-19 infection incidence level. Quite surprisingly, however, more deaths in the nursing home seems to reduce the gap in the tolerable infection incidence between Republican and Democrat governors, as illustrated by estimates in Columns (2) to (4) of Table \ref{tab:lambdaesti}. Governors from different parties therefore tend to converge in their policies when faced with a high death count. The estimation results also suggest that the gender of the governor determines the tolerable infection incidence level, with this level being higher in female governors by about 7\%. Moreover, being located in the South increases the tolerable infection incidence level by 5\%.

Our analysis suggests that ideological reasons and the severity of the epidemic impact the choice of the tolerable infection incidence. The U.S. state governor's gender and party affiliation and the state's geographical location are essential determinants of the tolerable infection incidence level.

\section{Policy Implications and Concluding Remarks} \label{sec:implicationsandconlusion}

This paper poses the problem of finding the optimal lockdown and reopening policy during a pandemic like COVID-19, for a social planner who \textit{prioritizes} health over short-term economic gains. Agents are connected through a \textit{weighted} network of contacts, and the planner's objective is to determine the policy that contains the spread of infection below a \textit{tolerable} incidence level and that maximizes the present discounted value of real income, in that order of priority. We show theoretically that the planner's problem has a unique solution. The optimal policy depends both on the configuration of the contact network and the tolerable infection incidence level. Simulation-based comparative statics analyses highlight the crucial role of network structure in infection spread and quantify the tradeoff between the tolerable infection level and human losses on the one hand, and the economic losses due to the pandemic on the other hand. The simulation exercises also show how different measures of network centrality correlate with the likelihood of being sent into lockdown and how that correlation varies with the tolerable infection incidence level. 

 We use unique data on the networks of nursing and long-term homes in the United States and from several other sources to calibrate our model. We uncover the tolerable COVID-19 infection incidence level, $\lambda$, for 26 U.S. states. A higher value of $\lambda$ indicates a preference for ``laissez-faire", short-term economic gains are maximized at the expense of population health. Using this parameter to test the model's predictions, we find that a ``laissez-faire" pandemic policy is associated with an increased number of deaths in nursing homes, while increasing U.S. state GDP growth. We also find significant interactions between $\lambda$ and other important variables. In particular, we find that ``laissez-faire" is more harmful to nursing homes that more peripheral in the networks compared to those that are more central. Given that network centrality is associated with more deaths, this means that ``laissez-faire" tends to decrease the importance of this network variable by increasing the vulnerability of peripheral nodes. ``Laissez-faire" is also more harmful to nursing that are located in deprived counties, and to those that work for a profit. This latter finding is important in the sense that our theoretical model is more valid for organizations that seek to maximize economic profit. Finally, in an attempt to validate our calibration exercise from external information, we examine the association between political variables and $\lambda$, finding that U.S. states with a Republican governor, southern states, and states with a female governor are more likely to tolerate the pandemic. However, the behaviors of Republican and Democratic governors tend to converge when faced with a high death count in nursing homes.\\

 \textbf{Funding:} Pongou  acknowledges  financial  support  from  the  SSHRC’s  Partnership  Engage  Grants  COVID-19  Special  Initiative.\\
 
\textbf{Conflict of Interest:} The authors declare that they have no conflict of interest.

\bibliographystyle{apalike2}
\bibliography{covid}
\newpage

\end{document}